\documentclass[manuscript]{aastex631}

\usepackage{newtxtext}  
\usepackage[T1]{fontenc}
\usepackage{multirow}
\usepackage[varvw]{newtxmath}

\newcommand{\Msun}{$M_\odot$}

\newcommand{\rgc}{$r_{\rm gc}$}
\newcommand{\rgce}{r_{\rm gc}}

\newcommand{\Msunpc}{$M_\odot / {\rm pc}^3$}

\newcommand{\kms}{km$\cdot$s$^{-1}$}

\newcommand{\vR}{v_{R}}
\newcommand{\vz}{v_{z}}
\newcommand{\vp}{v_{\phi}}
\newcommand{\sR}{\sigma_{R}}
\newcommand{\sz}{\sigma_{z}}
\newcommand{\sph}{\sigma_{\phi}}
\newcommand{\cRz}{\sigma_{Rz}^2}
\newcommand{\vrs}{v_{r}}
\newcommand{\vt}{v_{\theta}}
\newcommand{\uR}{u_{R}}
\newcommand{\uz}{u_{z}}
\newcommand{\up}{u_{\phi}}
\newcommand{\urs}{u_{r}}
\newcommand{\ut}{u_{\theta}}

\shorttitle{AASTeX v6.3.1 Sample article}
\shortauthors{Zhang et al.}
\graphicspath{{./}{figures/}}

\begin{document}

\title{The Shape and Mass of the Galactic Dark Matter Halo from the Axisymmetric Jeans Model}

\correspondingauthor{Xiang-Xiang Xue}
\email{xuexx@nao.cas.cn}


\author[0000-0001-7080-0618]{Lan Zhang}
\affiliation{National Astronomical Observatories, Chinese Academy of Sciences, Beijing 100101, People’s Republic of China}

\author[0000-0002-0642-5689]{Xiang-Xiang Xue}
\affiliation{National Astronomical Observatories, Chinese Academy of Sciences, Beijing 100101, People’s Republic of China}
\affiliation{Institute for Frontiers in Astronomy and Astrophysics, Beijing Normal University, Beijing 102206, People’s Republic of China}

\author[0000-0002-8005-0870]{Ling Zhu}
\affiliation{Shanghai Astronomical Observatory, Chinese Academy of Sciences, Shanghai 200030, People’s Republic of China}

\author[0009-0008-1319-1084]{Ruizhi Zhang}
\affiliation{National Astronomical Observatories, Chinese Academy of Sciences, Beijing 100101, People’s Republic of China}
\affiliation{School of Astronomy and Space Science, University of Chinese Academy of Sciences, Beijing 101408, People’s Republic of China}

\author[0000-0003-1972-0086]{Chengqun Yang}
\affiliation{School of Physics and Optoelectronic Engineering, Hainan University, Haikou 570228, People’s Republic of China}
\affiliation{Shanghai Astronomical Observatory, Chinese Academy of Sciences, Shanghai 200030, People’s Republic of China}

\author[0000-0001-8382-6323]{Shi Shao}
\affiliation{National Astronomical Observatories, Chinese Academy of Sciences, Beijing 100101, People’s Republic of China}

\author[0009-0003-7901-5327]{Jiang Chang}
\affiliation{Purple Mountain Observatory, Chinese Academy of Sciences, Nanjing 210034, People’s Republic of China}

\author{Feilu Wang}
\affiliation{National Astronomical Observatories, Chinese Academy of Sciences, Beijing 100101, People’s Republic of China}
\affiliation{School of Astronomy and Space Science, University of Chinese Academy of Sciences, Beijing 101408, People’s Republic of China}

\author[0000-0003-3347-7596]{Hao Tian}
\affiliation{National Astronomical Observatories, Chinese Academy of Sciences, Beijing 100101, People’s Republic of China}

\author[0000-0002-8980-945X]{Gang Zhao}
\affiliation{National Astronomical Observatories, Chinese Academy of Sciences, Beijing 100101, People’s Republic of China}
\affiliation{School of Astronomy and Space Science, University of Chinese Academy of Sciences, Beijing 101408, People’s Republic of China}

\author[0000-0002-1802-6917]{Chao Liu}
\affiliation{School of Astronomy and Space Science, University of Chinese Academy of Sciences, Beijing 101408, People’s Republic of China}
\affiliation{Institute for Frontiers in Astronomy and Astrophysics, Beijing Normal University, Beijing 102206, People’s Republic of China}
\affiliation{National Astronomical Observatories, Chinese Academy of Sciences, Beijing 100101, People’s Republic of China}



\begin{abstract}

We explore the density profile, shape, and virial mass of the Milky Way's dark matter halo using K giants (KG) from LAMOST and SDSS/SEGUE, as well as blue horizontal branch (BHB) stars from SDSS. Incorporating Gaia DR3 proper motions, we first investigate the velocity ellipsoid distribution within the $(R, |z|)$ space. The ellipsoids projected onto the $(v_R, v_z)$ plane exhibit near-spherical alignment.
We then probe the underlying dark matter distribution using the axisymmetric Jeans equations with multi-Gaussian
expansion (MGE) and the spherically aligned Jeans anisotropic modelling (JAM${\rm sph}$), allowing for different flattened dark matter density models. For each model, we apply two fitting approaches: fitting the KGs and BHBs separately or fit them simultaneously as two dynamical tracers in one gravitational potential. We find consistent results on the dark matter density profiles, $r_{200}$, and $M_{200}$ within a 1-$\sigma$ confidence region for models constrained by KGs, BHBs, and both.
We find the strongest consistency between KGs and BHBs in constraining dark matter profiles for models incorporating radially varying halo flattening ($q(r_{\rm gc})$), which suggests the Milky Way's dark matter halo shape evolves with Galactocentric distance (\rgc). Specifically, the halo flattening parameter $q_h$ decreases within $\rgce < 20$~kpc and increases for $\rgce > 20$~kpc. 
In this model, $M_{\rm tot} (< 60~{\rm kpc}) = 0.533^{+0.061}_{-0.054} \times 10^{12}$~\Msun, $r_{200}$ is $188\pm15$~kpc, with $M_{200}$ estimated at $0.820^{+0.210}_{-0.186} \times 10^{12} M_{\odot}$.
\end{abstract}



\section{Introduction}
\label{sec:intro} 
The mass of a galaxy's dark matter halo is a key in undstanding its intrinsic properties and significantly influences how the galaxy interacts with its surrounding environment, driving the formation and evolution of both the galaxy and the structure of the universe. 
The $\Lambda$-Cold Dark Matter \cite[$\Lambda$CDM][]{dode96} model has been successful in describing large-scale structures of the universe. However, challenges arise on smaller scales, particularly when comparing the model to the Milky Way (hereafter MW), e.g., the ``Missing Satellites'' problem \citep{kly99} and ``Too Big to Fail'' problem \citep{boly11}.
Determining the dark matter mass distribution in the MW accurately is crucial to addressing these issues.

Previous studies have estimated the total dark matter mass of the MW in the range of $\sim 0.5 - 2.5 \times 10^{12}$~\Msun. The uncertainty is due to the adopting of different model assumptions, tracers, and methodologies \citep[see Fig.~1, ][and references therein]{wang20}. 
Very recently, \citet{ou24} constructed the circular velocity curve of the MW out to $\sim$ 30~kpc,
and estimated a rather small dark matter halo mass $1.81^{+0.06}_{-0.05} \times 10^{11}$ \Msun with a cored Einasto dark matter density profile.
Despite these efforts, the assumption of a spherical halo has been increasingly questioned, as evidence suggests that the MW’s stellar halo is more oblate. While a spherical halo is often assumed for simplicity in theoretical and dynamical studies, facilitating the use of the spherical Jeans equation \citep[][hereafter B22]{bird22}, although this assumption is inconsistent with the halo's apparent ellipsoidal nature. Studies of various stellar populations, such as main sequence stars, RR Lyrae variables, blue horizontal branch stars (BHB), and K giants (KG), have revealed significant differences in the stellar density distributions and flattening \citep{sesar10, deason15, xue15, thomas18}. 

Until now, there is no definitive consensus on the shape of the Galactic dark matter halo. Some investigations suggest an oblate halo ($q_h < 1.$), for instance, $q_{\rm DM} = 0.7 \pm 0.1$ in \citet{loeb14} by using SDSS halo stars the heliocentric distances exceeding $\sim$ 10~kpc and galactocentric distances exceeding $\sim$ 20~kpc. Very recently, \citet{huang24} measured a $0.84 < q_h < 0.96$ of the Galactic dark matter halo by a retrograde precessing Galactic disk warp. Conversely, other studies propose a more spherical or even prolate halo ($q_h \ge 1.$). With RR Lyra between 5 and 20 kpc, \citet{wegg19} derived a nearly spherical dark matter halo, e.g, $q_{h} = 1.0 \pm 0.09$. \citet{posti19} derived a prolate dark matter halo ($q_{h} = 1.30 \pm 0.25$) with 75 globular clusters. These discrepancies are attributed to variations in tracers, methodologies, and spatial coverage, resulting in divergent conclusions about the halo’s geometry.
 A more precise understanding of the dark matter halo’s shape is critical, as it directly informs models of its mass distribution and offers deeper insights into the Milky Way’s formation and evolution.

Furthermore, most estimates of the MW's dark matter mass rely on a single stellar population as a tracer. Combining multiple stellar populations offers the potential to enhance constraints on the MW’s gravitational potential and improve the precision of mass estimates. However, this approach requires a more sophisticated and flexible dynamical model, as different populations exhibit varying spatial distributions and kinematic properties.

In this study, we adopt an axisymmetric model for the MW’s dark matter halo and stellar number density profiles, which better reflects observed properties. By solving the axisymmetric Jeans equation for multiple tracers, such as KG and BHB stars, we aim to more effectively constrain the total mass of the MW and the shape of its dark matter halo.

This paper is organized as follows. The data used in  this work are introduced in
Sec.~\ref{sec:data}. The method is described in Sec.~\ref{sec:method}. The results are shown and discussed in
Sec.~\ref{sec:results}, with a conclusion in Sec.~\ref{sec:conclusion}.


\section{Data}
\label{sec:data}

\subsection{KG \& BHB samples}
\label{subsec:tracers}
To probe the Milky Way's dark matter halo morphology and mass profile out to $\rgce \geq 40$~kpc, we use KG and BHB as tracers, which have complete six dimension phase-space coordinates extending to 120~kpc.

KGs used in the present work are from halo catalogs of \citet{zhang23} and \citet{xue14}, in which
samples are selected from the LAMOST DR8 and SDSS/SEGUE DR9 \citep{ahn12}, respectively. 
LAMOST is a quasi-meridian reflecting Schmidt telescope with an effective aperture of
4 meters and 4000 optical fibers \citep{cui12}, from which numerous low-resolution
($R\sim1800$) spectra covering a wavelength range of $3700 < \lambda < 9000$~{\AA}
of stars with $r < 19$ can be obtained simultaneously at one exposure \citep{zhao06, zhao12},
and SDSS and its extensions use a dedicated 2.5 m telescope \citep{gunn06}
to obtain $ugriz$ imaging \citep{fuku96, gunn98, york00, stou02, pier03, eise11, blan17}.
SEGUE, one of the key projects executed during SDSS-II and SDSS-III, obtained some 360,000
spectra of stars in the Galaxy, selected to explore the nature of stellar populations from 0.5~kpc
to 100~kpc \citep{yanny09, rock12}.

The catalogs of KGs include distances determined using the Bayesian approach by \citet{xue14}.
The KGs extend up to $\sim$ 120 kpc, with a mean relative distance error of 14.5\% and a typical radial velocity error of around 7~\kms.
While the BHBs, taken from \citet{xue11}, consists of stars selected from SDSS/SEGUE DR8 \citep{aiha11}. 
To ensure that the KGs consist of chemically and kinematically confirmed halo tracers, only stars locate at $|z| > 5.0$~kpc, and ${\rm [Fe/H]} < -1.0$ if 2.0~kpc $< |z| <$~5.0 kpc
are adopted.
The distances for BHBs were
derived based on magnitudes and colors following the method in \citet{xue08}. This BHB sample provides precise distance estimates, with an accuracy of about 5\%,
and covers a range of up to $\sim$ 80 kpc, with about 90\% of the sample lying between 5 kpc - 40 kpc. Moreover, the radial velocity errors for the BHBs range from 5 to 20~\kms

For both of the tracers used in this work, we clean substructures to get at a smooth halo by applying the friends-of-friends
(hereafter FoF) algorithm in the integrals of motion (hereafter IoM) space \citet[][Xue et al., in preparation]{xue11}. This algorithm
has been well validated. For instance, \citet{wang22} used the same method to identify three previously unrecognized substructures
that may be associated with globular clusters NGC 5272, NGC 6656, and NGC 5024. Additionally, the FoF algorithm successfully identified known substructures,
including the Sagittarius stream \citep{new02, maj03}, the Gaia-Enceladus-Sausage (GES) \citep{mye18, bel18}, the Sequoia \citep{mye19}, and the Helmi streams \citep{hel99}.

\begin{figure}
	\gridline{\fig{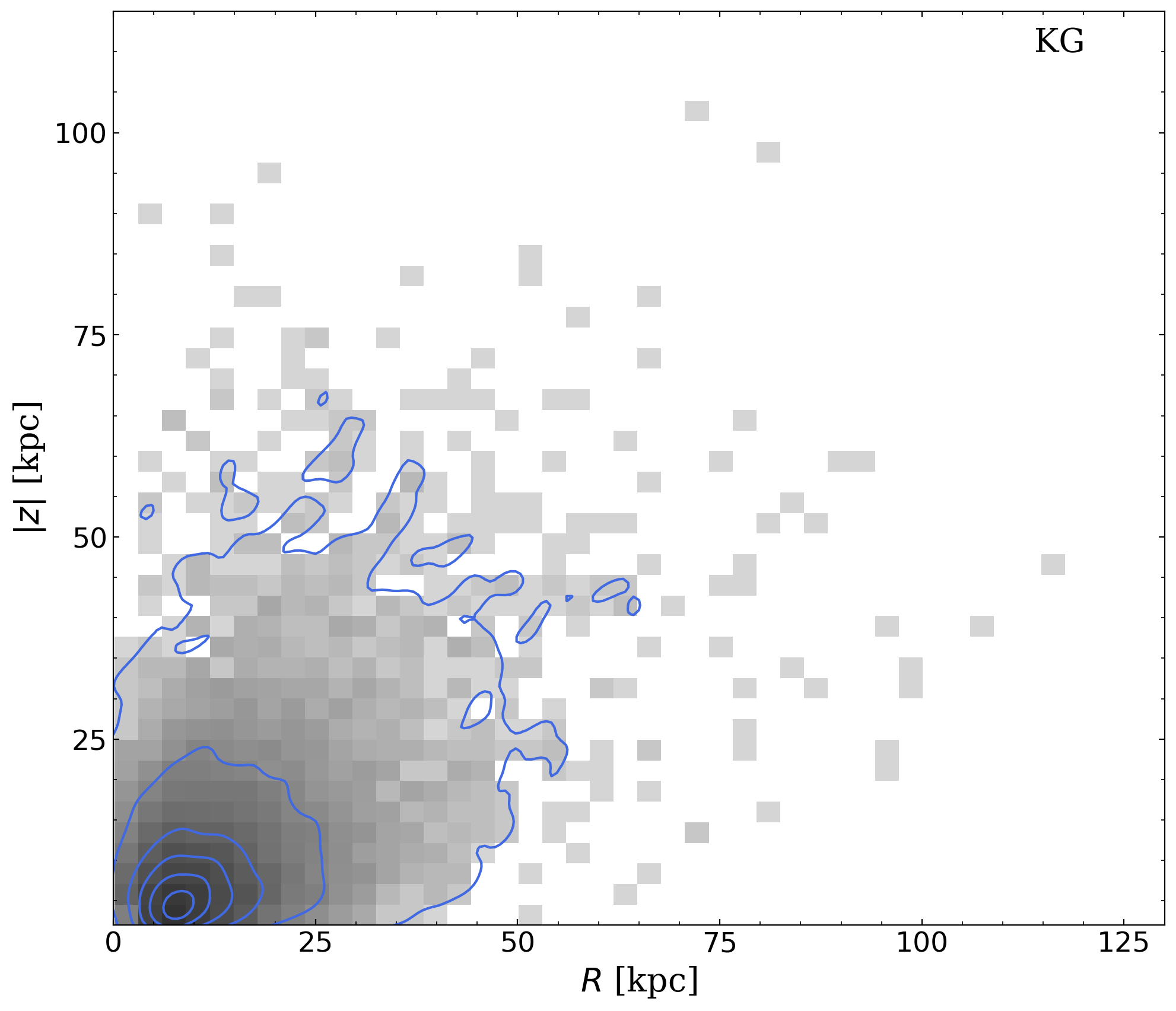}{0.4\textwidth}{(a)}}
	\gridline{\fig{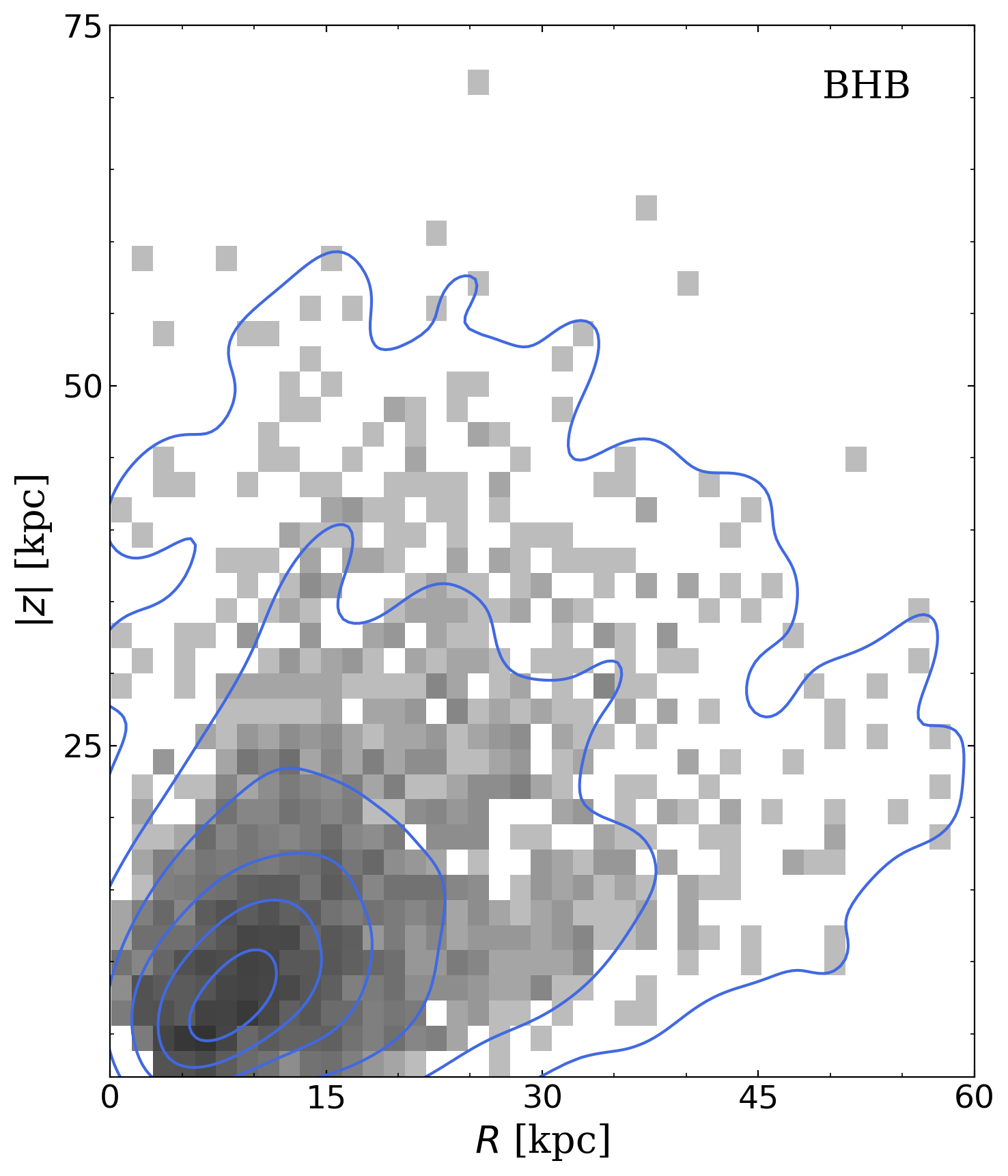}{0.4\textwidth}{(b)}}
	\gridline{\fig{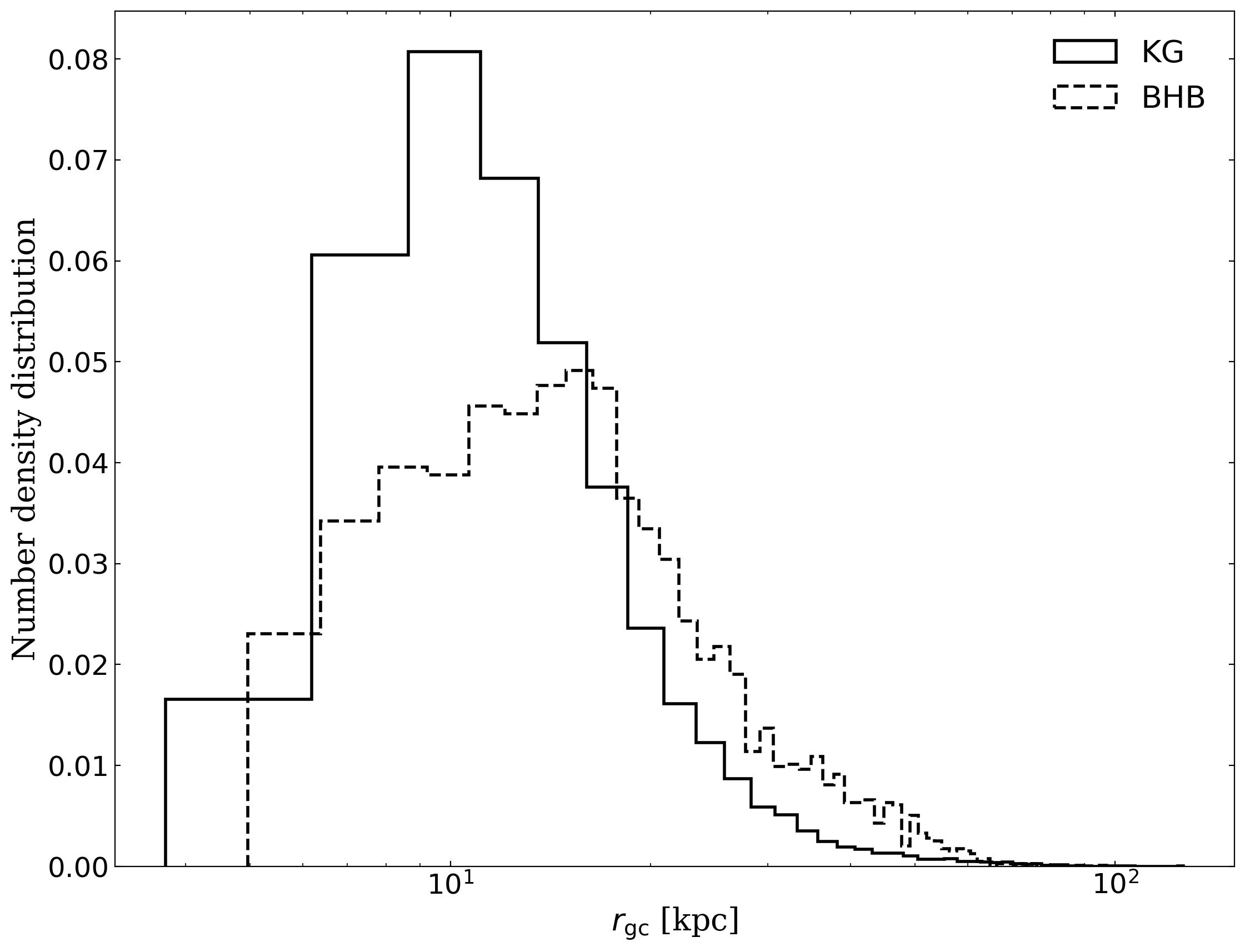}{0.4\textwidth}{(c)}}
    	\caption{Number density distributions of tracers used in the present study. The plots from top to bottom are: (a) the density distribution of KG
	in $(R, |z|)$ space; (b) similar to (a) but for BHB; (c) the PDF along with the galactocentric radius $r_{\rm gc}$, black solid and dashed lines represent
	KG and BHB, respectively.}
\label{fig:num_den}
\end{figure}

We performed FoF clustering on the full sample of 54,762 KGs and 4,589 BHBs in the integrals-of-motion space (i.e. energy and angular momentum),
removing substructures with $\geq 5$ member stars. This yielded cleaned smooth halo tracers, e.g., 21,839 KGs and 2,767 BHBs. The distributions of the tracers in $R-z$ space and and along with the galactocentric distance $\rgce$ are shown in Fig.~\ref{fig:num_den}.
It can be seen that our samples can trace as far as $r_{\rm gc} \sim 120$~kpc. 
Although the sample size of BHBs is smaller and their distance coverage
is less extensive than that of KGs, the number density of BHB stars is larger at $\rgce > 15$~kpc, making them more
effective tracers of the outer halo.

\subsection{Kinematics}
\label{subsec:kine}
With proper motions extracted from the Gaia Data Release 3
\citep[DR3;][]{gaia16, gaia21}, we first calculated the rectangular velocity
components to the Galactic center (GC) by adopting a value of 238~\kms
\citep{scho12} for the local standard of rest ($V_{\rm LSR}$) and a solar motio
of (+9.58, 10.52, 7.01)~\kms \citep{tian15} in ($\mathit{U}$, $\mathit{V}$,
$\mathit{W}$), which are defined in a right-handed Galactic system with $\mathit{U}$
pointing toward the GC, $\mathit{V}$ in the direction of rotation, and $\mathit{W}$
toward the north Galactic pole \citep{dehn98}. The kinematic parameters including
distance and velocities are then transfer to the Galactic cylindrical coordinates:
\begin{equation}
\begin{aligned}
       & \phi = \arctan \left[\frac{D\cos(b)\sin(l)}{R_{\sun} - D\cos(b)\cos(l)} \right]\\
       & \boldsymbol{v}_{\rm gc} = {\rm \Pi}\hat{\boldsymbol{e}}_R + {\rm \Theta}\hat{\boldsymbol{e}}_{\phi}+ {\rm Z}\hat{\boldsymbol{e}}_z\\
\end{aligned}
\end{equation}
where $(l, b)$ are observed Galactic longitude and latitude, $R_{\odot} = 8.34$~kpc
\citep{reid14} is the adopted distance to the GC, $\Pi = \vR \equiv \dot{R}$,
$\Theta = \vp \equiv R \dot{\phi}$, and $Z = \vz \equiv \dot{z}$. 

\begin{figure} [ht!]
\centering
\includegraphics[width=\textwidth7]{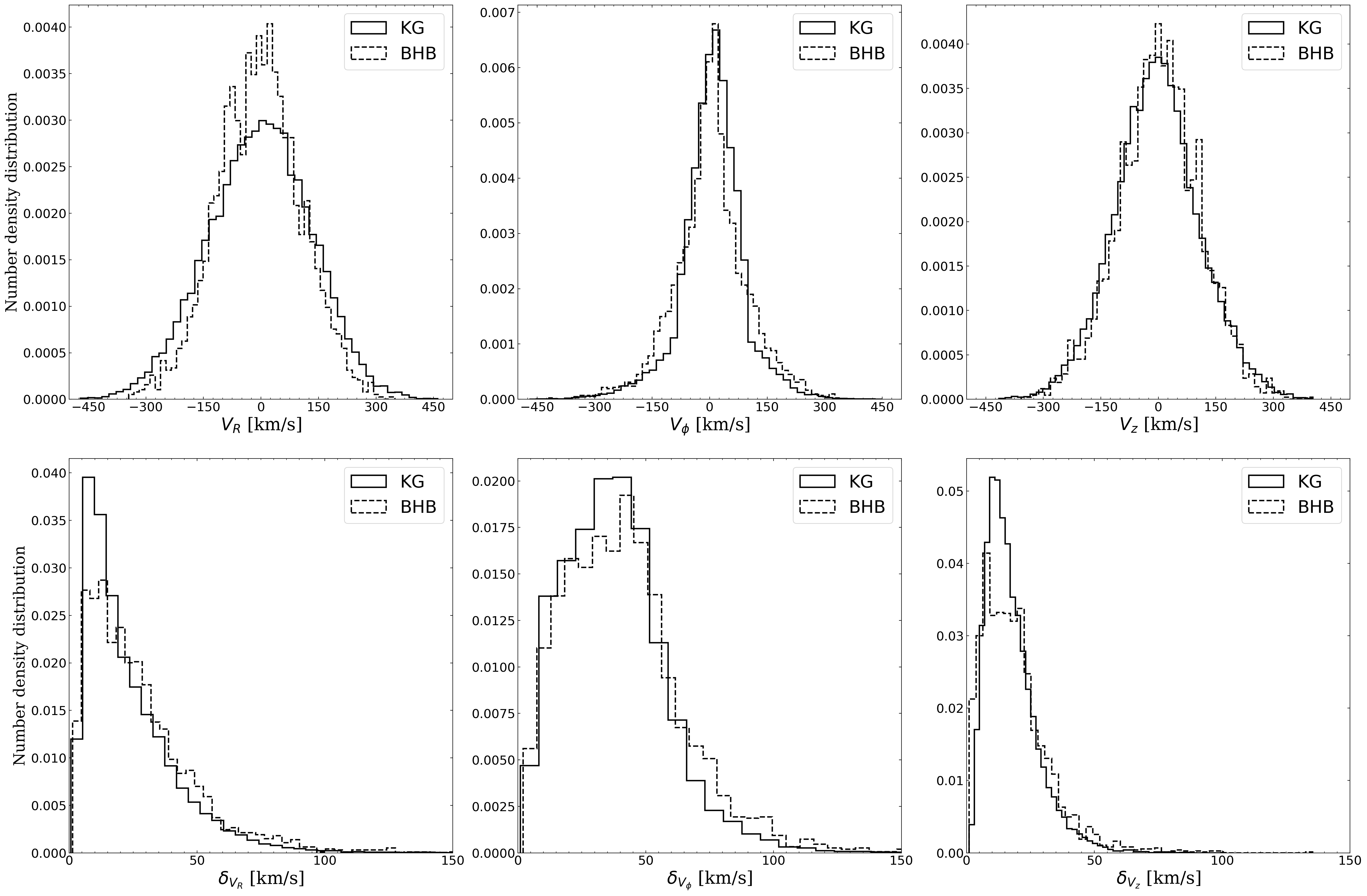}
\caption{The number density distributions of velocities and corresponded errors of tracers. Black solid and dotted lines represent the mean velocity of
             KG and BHB, respectively.}
	     \label{fig:obs_vel_PDF}
\end{figure}

The error matrix of velocity components, $\boldsymbol{\Delta}$, are estimated
by Monte Carlo sampling with assuming a multivariate Gaussian distribution.
For each star, 1000 sets of variables used in velocity calculation are generated
from a six-dimension multivariate Gaussian distribution, $\mathcal{N}(\boldsymbol{\mu},
\boldsymbol{\Sigma_{\delta}})$, where $\boldsymbol{\mu} = [{\rm ra}, {\rm dec}, \mu_{\alpha},
\mu_{\delta}, D, v_{\rm rad}]$ are observed values, and $\boldsymbol{\Sigma_{\delta}}$
is the covariance matrix calculated by observation errors and correlation coefficients
provide by Gaia DR3 catalog. Thus, a three-dimensional velocity distribution for each
star is derived. The means and covariance of the distribution are adopted as calculated
velocity values $\boldsymbol{v} = [\vR, \vp, \vz]$ and error matrix $\boldsymbol{\Delta}$, respectively.
Distributions of the velocities and their corresponded errors are shown in Fig.~\ref{fig:obs_vel_PDF}. It can be seen that the substructures are well removed.
The typical velocity errors $\delta$ in $\vR$, $\vp$, and $\vz$ are 18~\kms, 35~\kms, and 15~\kms, respectively. 

\begin{figure} [ht!]
\centering
\includegraphics[scale=0.37]{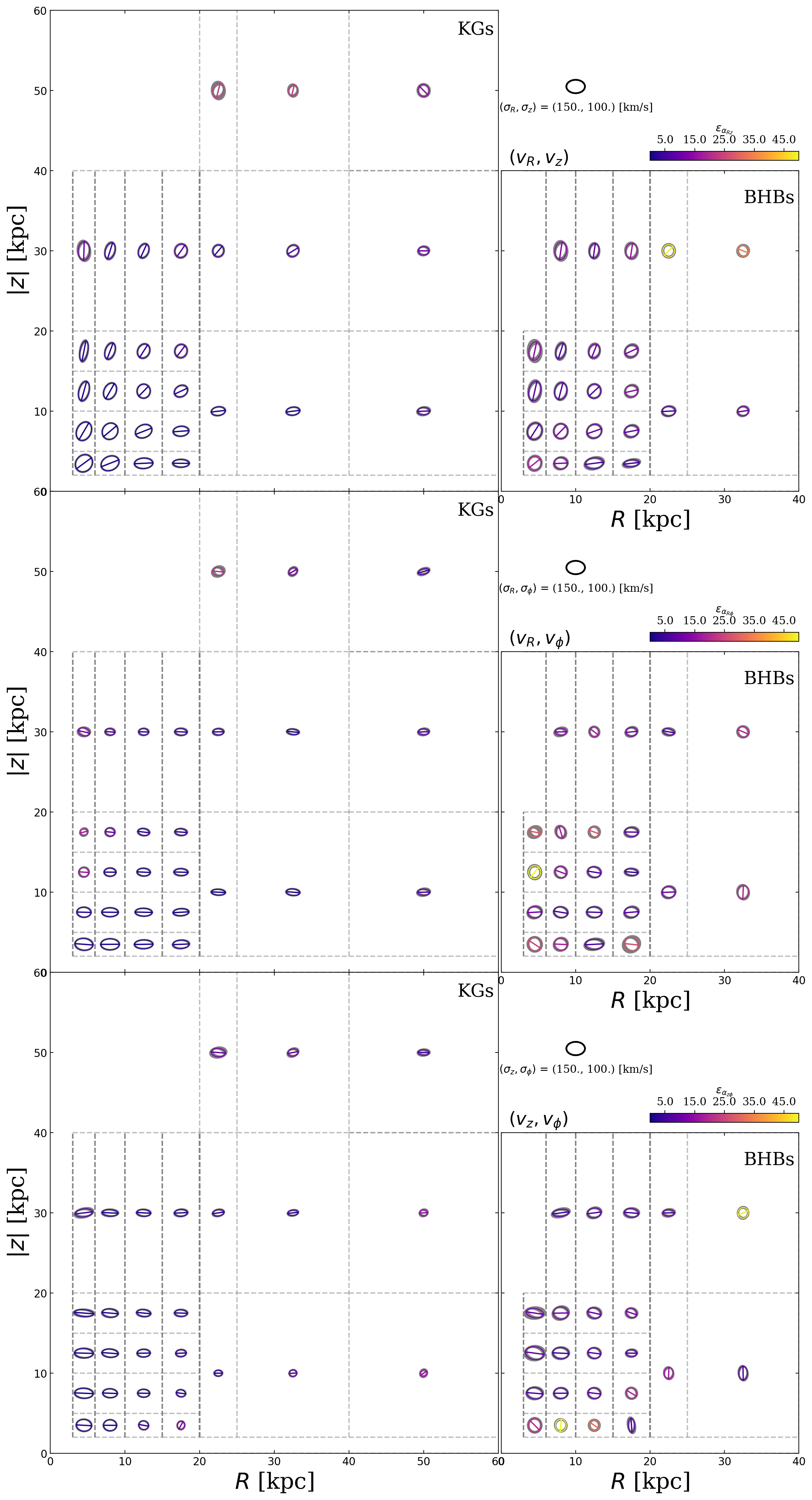}
\caption{Velocity ellipsoids derived from observations for each tracer. From the top to the bottom,
	      the plots show the ellipsoids projected in $(\vR, \vz)$, $(\vR, \vp$), and ($\vz, \vp)$ spaces,
	      respectively. In each space, the color ellipses are the best calculated values from the observations, 
	      while the grey shades are corresponded $1-\sigma$ error regions. Thick lines indicate the directions of each velocity ellipse.
	      The color code indicate the error of the tilt angle of the velocity ellipsoids.
	      The first column are the results of the KG, and the second column represents the ones of the BHB.}
	     \label{fig:obs_ta_all}
\end{figure}

In this coordinate system, the observed velocity distributions of KGs and BHBs are analyzed using a grid of
$R - |z|$ cells. For the KGs, the radial divisions are defined by $R-$edges at $[3, 6, 10, 15, 20, 25, 40, 130]$~kpc. 
The vertical divisions vary depending on the radial range: for $R \leq 20$~kpc, $|z|-$ edges are defined as
$[2, 5, 10, 15, 20, 40, 60]$~kpc, while for for $R > 20$~kpc, $|z|-$ edges are set to $[0, 20, 40, 60]$~kpc.
The BHB grid follows a similar structure but is truncated at $R = |z| = 60$~kpc.
In 
each $R - |z|$ cell, the intrinsic velocity distribution of the stars is assumed as
$\mathcal{L}(\boldsymbol{v})$  in ($\vR$, $\vp$, $\vz$) space.
Each stellar velocity $\boldsymbol{v_{k}}$ in this cell is the product of $\mathcal{L}(\boldsymbol{v})$ convolved with a delta function which is broadened by the error matrix $\boldsymbol{\Delta_k}$, and integrated over all velocities. For all $N$ stars in the cell, the likelihood can be defined by
\begin{equation}
\begin{aligned}
\label{eq:v_like}
\mathscr{L}(\bar{\boldsymbol{v}}, \boldsymbol{\Sigma_{\rm I}}) = & \prod_{k=1}^{N}\int_{-\infty}^{+\infty}\,\mathcal{L}(\boldsymbol{v})  \frac{1}{\sqrt{(2\pi)^3\det(\boldsymbol{\Delta_k})}} \times \\
&  \exp\left[-\frac{1}{2}(\boldsymbol{v_k} - \boldsymbol{v})^{\rm T} \boldsymbol{\Delta}^{-1}_k (\boldsymbol{v_k} - \boldsymbol{v})\right] \, d \boldsymbol{v},
\end{aligned}
\end{equation}
where $\mathscr{L}$ is a function of mean velocity $\bar{\boldsymbol{v}}$, and velocity dispersion tensor
$\boldsymbol{\Sigma_{\rm I}}$., and the velocity distribution $\mathcal{L}(v_z)$ is described by a parameterized normal distribution.
All the free parameters in each $R - |z|$ cell  which $N > 50$ are considered, $\bar{\boldsymbol{v}}$ and $\boldsymbol{\Sigma_{{\rm I}}}$ are predicted by a Markov Chain Monte Carlo (MCMC) analysis based on the Metropolis-Hasting
method \citep{pres07} and an affine-invariant ensemble sampling algorithm \citep{good10},
{\it emcee}, which was developed by \citet{emcee}. The derived velocity ellipsoid in each $R-|z|$ cell, along with the corresponding errors
which calculated from the samplings of the last 200 steps of post-burn-in iterations in the MCMC chain is shown in Fig.~\ref{fig:obs_ta_all}.
For the KG, the bin ranging from 40 to 60~kpc includes all samples with $R > 40$~kpc, extending out to as far as 130~kpc.
Similarly, for the BHB, the bin ranging from 25 to 40~kpc includes all samples with $R > 25$~kpc.
In most $R-|z|$ cells, the error-indicating shading almost coincides with the best-fit ellipse, indicating very small errors on velocity dispersions.
The velocity dispersion uncertainties exhibit radial progression from $\sim 5$~\kms ($\rgce < 20$~kpc, $N > 500$ stars per bin) to $\sim 8$~\kms
($\rgce > 40$~kpc, $N \sim 50$ stars per bin) with localized enhancements at large radii due to small sample size.
It can be seen that the observed velocity ellipsoids projected onto the $(\vR, \vz)$ plane present a nearly spherically
alignment in the meridional plane $(R, z)$, while the velocity ellipses in $(\vR, \vp)$ and $(\vz, \vp)$ are oriented with the cylindrical coordinate system.

\section{Method}
\label{sec:method}
In this work, we adopt the axisymmetric Jeans model to constrain the underlying gravitational potential of the MW.
Using the observed tracer density profile $\nu_{\star}$ of two different types of tracer stars, BHB and KGs,
and a set of parameterized gravitational potential models $\Phi$, which describe the volume density of baryonic
matter $\rho_{\rm BM}$ and dark matter halo $\rho_{\rm DM}$, we solve the axisymmetric Jeans equations and to predict the velocity
dispersion tensor $\sigma_{ij}^2(\boldsymbol{r})$ \citep[$\sigma_{ij}^2(\boldsymbol{r}) \equiv \overline{v_i v_j} - \overline{v_i}~\overline{v_j}$,][]{binney08}
at the position $\boldsymbol{r}$ in a set of orthogonal axes $\hat{\boldsymbol{e}}_i$ ($\hat{\boldsymbol{e}}_j$)
 for each $\Phi$, and compared it with the observed value from the tracers to find the best potential parameters, 
 and hence to determine the dark matter density distribution of the MW and hence estimate its total mass.


\subsection{Axisymmetric Jeans equations}
\label{subsec: jeans}
Typically, the velocity dispersion components $\sigma_{ij}^2 (i  \neq j)$ are assumed to 0. to solve the axisymmetric Jeans equation, and
 directly solving the axisymmetric Jeans equation in cylindrical coordinates aligns well with the intuitive understanding.
However, from Fig.~\ref{fig:obs_ta_all}, it can be seen that velocity ellipses in the $(\vR, \vz)$ plane are
nearly spherically aligned, e.g.,  $\sigma_{Rz}^2 > 0$.  Instead, it is more reasonable to solve the anisotropic Jeans equation in spherical coordinates to match the
velocity ellipsoid alignments of the observed data,
\begin{subequations}
	\begin{align}
	\label{eq:jeans_eq}
		& \frac{\partial(\nu_{\star} \overline{\vrs^2})}{\partial r} + \nu_{\star} \frac{2\overline{\vrs^2} - \overline{\vt^2} - \overline{\vp^2}}{r} + \nu_{\star} \frac{\partial \Phi}{\partial r} = 0, \\
		& \frac{\partial(\nu_{\star} \overline{\vt^2})}{\partial \theta} + \nu_{\star} \frac{\overline{\vt^2} - \overline{\vp^2}}{\tan \theta} + \nu_{\star} \frac{\partial \Phi}{\partial \theta} = 0,
\end{align}
\end{subequations}
with an assumption that the cross-terms of the second velocity moment tensor vanish. 
In this case, after solving the equations, we can obtain velocity dispersion tensors in cylindrical coordinates that are consistent with observations through coordinate transformations.
\begin{equation}
 \begin{bmatrix} 
  \overline{\vR^2} & \overline{\vR \vz} & 0. \\
  \overline{\vR \vz} & \overline{\vz^2} & 0. \\
  0. & 0. & \overline{\vp^2} \\
  \end{bmatrix}  = \boldsymbol{\rm Q} \cdot  \begin{bmatrix} 
  \overline{\vrs^2} & 0. & 0. \\
  0. & \overline{\vt^2} & 0. \\
  0.  & 0. & \overline{\vp^2} \\
  \end{bmatrix} \cdot  \boldsymbol{\rm Q^T}
\end{equation}
where
\begin{equation}
\boldsymbol{\rm Q} = \begin{bmatrix} 
  \sin \theta & \cos \theta & 0. \\
  \cos \theta & -\sin \theta & 0 .\\
  0.  & 0. & 1. \\
  \end{bmatrix}, 
\end{equation}
and the angle $\theta$ are measured from the symmetry $z$-axis.

\citet{capp08, capp20} presented a flexible solution for the anisotropic spherically aligned axisymmetric
Jeans equations, JAM$_{\rm sph}$, by performing a Multi-Gaussian Expansion \citep[MGE;][]{ems94, capp02} of the stellar number density $\nu$ and
mass density distribution $\rho$, that is,
\begin{equation}
\label{eq:nu_mge}
\nu_{\star}(R, z) =  \sum_{n=1}^{N} \nu_{0,n} \exp \left[-\frac{1}{2 \sigma_n^2} \left( R^2 + \frac{z^2}{q_n^2}\right)\right],
\end{equation}
and 
\begin{equation}
\label{eq:rho_mge}
\rho(R, z) =  \sum_{m=1}^{M} \rho_{0, m} \exp \left[-\frac{1}{2 \sigma_m^2} \left( R^2 + \frac{z^2}{q_m^2}\right)\right],
\end{equation}
where $\sigma_n$ and $\sigma_m$ are dispersions along the major axis $R$, $q_n$ and $q_m$ are intrinsic axial ratios of each individual component, and $N$ and $M$ are the total numbers of expansed Gaussians. The forms can also be converted to spherical coordinates using the transformation
\begin{equation}
\label{eq:trans}
	(R, z) = (r \cos{\theta}, r \sin{\theta}).
\end{equation}
In the solution, the author defined the axial ($\beta$) and tangential ($\gamma$) anisotropies as

\begin{subequations}
	\begin{align}
		\label{eq:bg_b}
		& \beta = 1. - \sigma^2_{\theta} / \sigma^2_{r}, \\
		\label{eq:bg_g}
		&  \gamma = 1. - \sigma^2_{\phi} / \sigma^2_{r}, \\
		\label{eq:vphi2}
		& \overline{\vp^2} = \overline{\vp}^2 + \sigma^2_{\phi}
	\end{align}
\end{subequations}

and assumed that $\overline{\vrs} = \overline{\vt} = 0.$ \kms. Therefore, the anisotropy of the MW
\begin{equation}
\label{eq:aniso}
	\beta' = 1. - \frac{\sigma^2_{\theta} + \sigma^2_{\phi}}{2\sigma^2_{r}} = \frac{1}{2} (\beta + \gamma)
\end{equation}

With the Poisson equation, which relates the potential to the mass density, the $\Phi$ generated by the density form of Eq.~\ref{eq:rho_mge} was given by \citet[Eq.~40,][and reference therein]{capp20}. Therefore, given the stellar number density and total mass density profiles, and using the MGE and JAM$_{\rm sph}$
solution described above, the components $[\sR, \sz, \sph, \cRz]$ of the velocity dispersion tensor 
can be derived. This method has been successfully applied by \citet{nits21} to conduct a detailed study of the mass distribution in the Galactic disk. 

To obtain the dark matter mass density distribution and hence to estimate the total mass of the MW, we adopt a parameterized form of the total mass density profile, which is assumed to be the sum of contributions from baryonic matter and from dark matter
\begin{equation}
\label{eq:rho_tot}
\rho_{\rm tot}(R, z | \boldsymbol{p})= \rho_{\rm BM} (R, z) + \rho_{\rm DM} (R, z |  \boldsymbol{p}),
\end{equation}
where $\boldsymbol{p}$ is a parameter set of any given form of dark matter density profile. Note that we did not parameterize the baryonic matter profile because our sample consists of halo stars far from the Galactic disk, primarily affected by the dark matter halo,  and have little constrains on the baryonic mass.

Then for a given tracer, we proceed as follows:
\begin{enumerate}
	\item[(1)]  The total mass density profile $\rho_{\rm tot} (R, z |  \boldsymbol{p})$ is modeled.
	\item[(2)] The given stellar number density profile $\nu(R, z)$ and $\rho_{\rm tot} (R, z |  \boldsymbol{p})$ are approximated using Eqs.~\ref{eq:nu_mge}
	             and \ref{eq:rho_mge}, respectively, with MGEFIT package \citep{capp02}\footnote{\url{https://pypi.org/project/mgefit/}}.
      	\item[(3)] The JAM$_{\rm sph}$ model is applied using the JAMPY package \citep{capp08, capp20} \footnote{\url{https://pypi.org/project/jampy/}} 
	              incorporating the MGE results and the unknown anisotropy parameters $\beta$ and $\gamma$. In this work, we do not explore the variance of
	              $\beta$ and $\gamma$ with \rgc, to reduce the number of fitting parameters.
	              This allows for solving the velocity dispersion tensor
	              $\boldsymbol{S}(R, z | \boldsymbol{p}, \beta, \gamma)$.
       	\item[(4)] For each $[\boldsymbol{p}, \beta, \gamma]$ of a given star, the dynamical probability
                     $P_{\rm dyn}(\boldsymbol{S}| \boldsymbol{p}, \beta, \gamma)$ is calculated.
	\item[(5)] An MCMC technique is employed to sample the likelihood of the data, given $[\boldsymbol{p}, \beta, \gamma]$ in order to find the best parameters of the dark matter density.
\end{enumerate}
Here the details are described step by step below.
 
\subsection{Mass density}
\label{subsec:gp}
The total mass density $\rho_{\rm tot}$ (Eq.~\ref{eq:rho_tot}) has contributions from both the baryonic components and the dark matter.
For baryonic matter of the MW, we following the assumption of \citet{watkins19}, it is composed of a nucleus,
bulge, and disk, e.g.,
\begin{equation}
\label{eq:eho_bm}
	\rho_{\rm BM} = \rho_{\rm nucleus} + \rho_{\rm bulge} + \rho_{\rm disk}.
\end{equation} 
The sets of nucleus and bulge are adopted from \citet{pw17}, in which both of the two components are characterized using the Herquist
potential, that is, a nucleus with mass of $M_{\rm nucleus} = 1.71 \times 10^{9}$~\Msun
and scale length of $l_{\rm nucleus} = 0.07$~kpc, and a bulge with mass of $M_{\rm bulge}
= 5.0 \times 10^{9}$~\Msun and scale length of $l_{\rm bulge} = 1.0$~kpc. We used
a sum of three Miyamoto-Nagai (3MN) disk potentials \citep{smith15} to approximate
the thick and thin disks. For the thin disk, its total mass of $M_{\rm thin} = 4.6 \times
10^{10}$~\Msun and radial exponential scale length $h_R = 2.2$~kpc are adopted from
\citet{bovy13}, and exponential scale height $h_z = 0.2$~kpc is taken from \citet{larsen03}.
While for the thick disk, $h_R = 3.8$~kpc, $h_z = 0.9$~kpc \citep{moni12}, and $M_{\rm thick}
= 4.0 \times 10^{9}$~\Msun, which is  8.6\% of the thin disc mass \citep{yoa06}.

For the dark matter halo, we adopt the density distribution of \citet[][hereafter NFW]{nfw97}
\begin{equation}
\label{eq:rho_dm}
\rho_{\rm DM} = \frac{\rho_s}{(r_{\rm eff} / r_s) (1 + r_{\rm eff} / r_s)^{2}}.
\end{equation}
In the axisymmetric case, $r_{\rm eff}^2 = x^2 + y^2 + (z/q_h)^2$, where $q_h$ is the flattening
 of the dark matter halo. Hence, there are two free halo parameters: the scale radius $r_s$, and the scale
density $\rho_s$. In this analysis, we aim to investigate the shape of the dark matter halo and the effects of different flattening forms
on its distribution and total mass. Therefore, we considered three different forms of $q_h$: a spherical dark matter halo,
$q_h$ as a free parameter, and $q_h$ as a step function along \rgc.

\subsection{Number density of tracers}
\label{subsec:nd}
In recent years, several studies have explored the stellar halo profile and made measurements of the number density of KG and BHB. Therefore, in this work, we will directly use the measurement results from the literatures.

X15 used SDSS/SEGUE-II KGs spanning $10 < \rgce < 80$ ~kpc to fit different number density profile models after correcting for selection function of SEGUE-II  and removing substructures. They found that a broken power law (BPL) with a constant flattening, $q$ ($r_q  \equiv \sqrt{R + (z / q)^2}$),  can fit the density profile of KGs, or alternatively, a single power law (SPL) with variable flattening ($q_{\rgce}$) can also fit the data well. With LAMOST DR3 data, \citet{xu18} reached similar conclusions, that is, the flattening increases with the increasing of $\rgce$, although their number density profile only covered up to 30 kpc.

\citet{thomas18} selected BHB stars with $15 < \rgce < 220$ ~kpc from the Canada–France Imaging Survey (CFIS) and, combined with Pan-STARRS 1 $griz$ photometry to understand the stellar halo profiles of the MW. They concluded that the outer stellar halo traced by the BHB stars is well modeled by a BPL with a constant flattening. Additionally, they provided fitting results for the other two profile models, i.e., an SPL with a constant flattening and an SPL with variable flattening. However, the results for the latter model presented a larger error range and higher Bayesian Information Criterion (BIC) value.

\begin{figure}
	\gridline{\fig{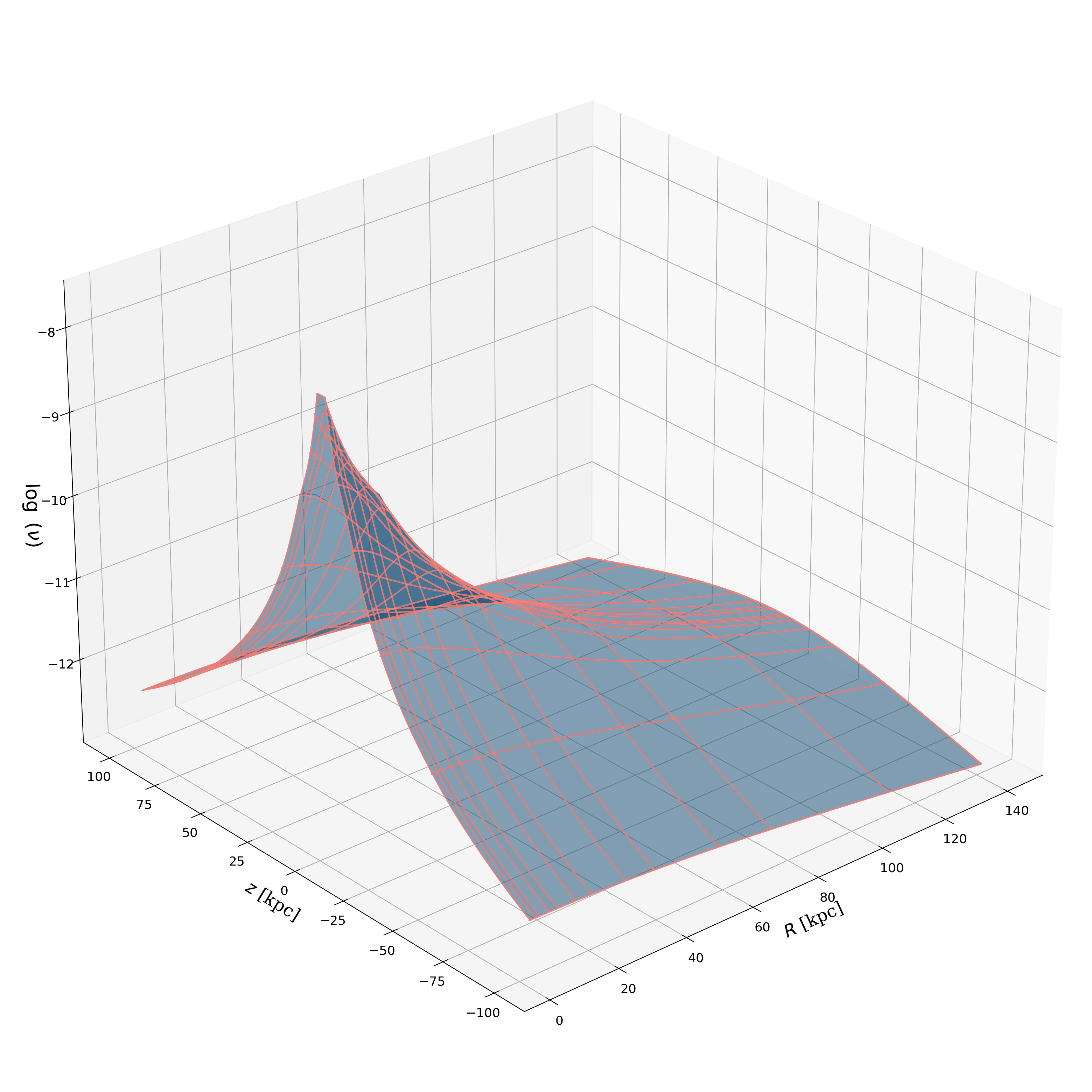}{0.4\textwidth}{(a) KG}
	             \fig{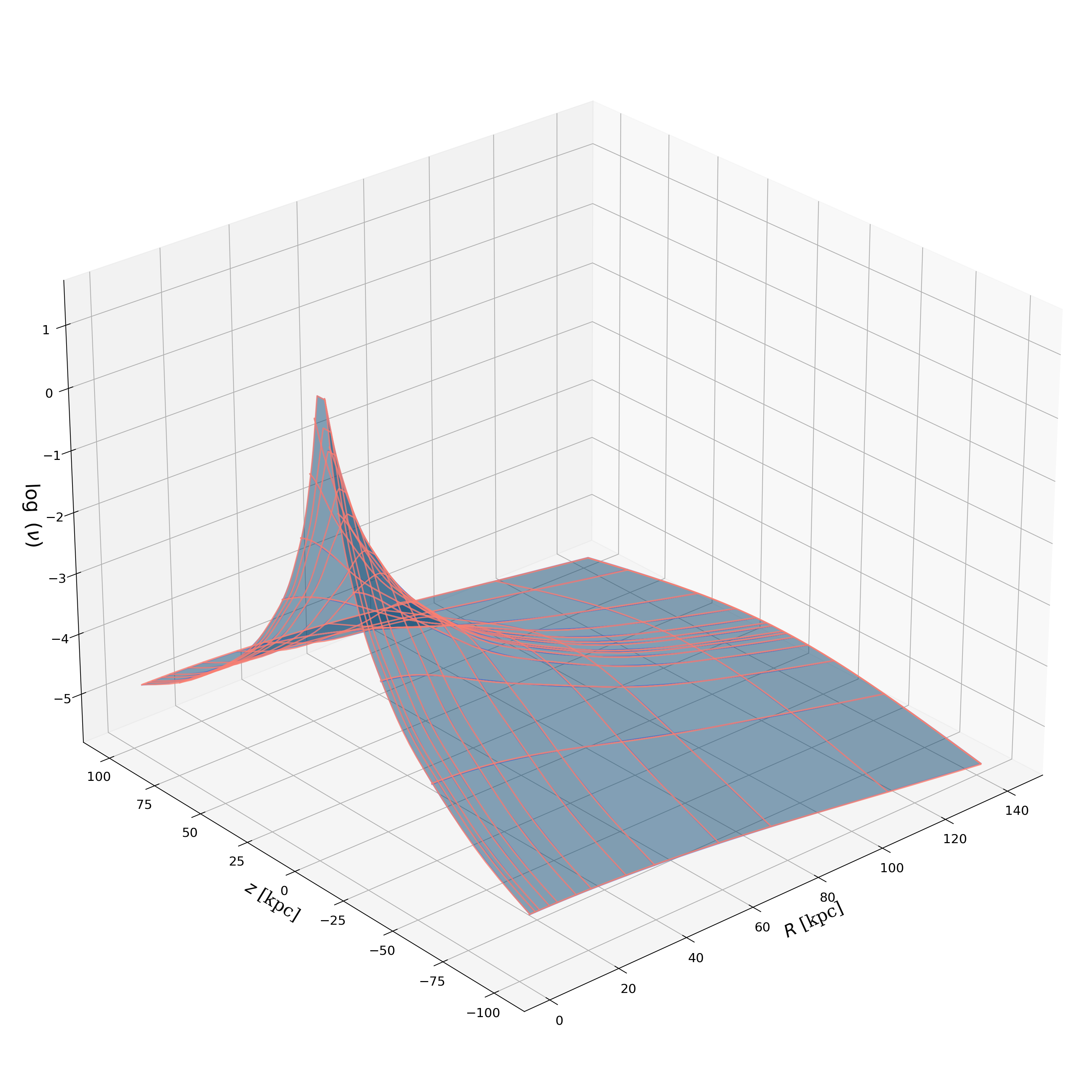}{0.4\textwidth}{(b) BHB}}         
    	\caption{MGE for number density profiles of the tracers. In the plots,
	             the blue shades are number density profiles from Eqs.~\ref{eq:bpl_kg} and \ref{eq:bpl_bhb}, and red wires are results from MGE fitting.}
\label{fig:num_den_mge}
\end{figure}

We tested the MGE fitting for the profiles listed in the aforementioned literatures. The results show that for both KG and BHB, the SPL profile is difficult to fit within
$\rgce < 5.$~kpc, with $\chi^2$ values higher than those for the BPL profiles. Additionally, to ensure consistency, the BPL profile for both of tracers are adopted in the present analysis,
\begin{itemize}
	\item[(1)] KG, fitting results is from X15,
		\begin{equation}
		\label{eq:bpl_kg}
			\nu_{\star}(\rgce) = \nu_0
			\begin{cases}
		 		r_q^{-\alpha_{\rm in}}, & r_q \leqslant r_{\rm break} \\
				r_{\rm break}^{(\alpha_{\rm out} - \alpha_{\rm in})} \times r_q^{-\alpha_{\rm in}},  & r_q > r_{\rm break},
			\end{cases}
	\end{equation}
	where $\alpha_{\rm in} = 2.1 \pm 0.3$, $\alpha_{\rm out} = 3.8 \pm 0.1$, $r_{\rm break} = 18.0 \pm 1$~kpc, and $q = 0.7 \pm 0.02$. $\nu_{0}$ is fix to 1 in our model, since it is a
	constant and can be eliminated in the Eqs.~\ref{eq:jeans_eq}.
	\item[(2)] BHB, fitting result is from \citet{thomas18},
		\begin{equation}
		\label{eq:bpl_bhb}
			\nu_{\star}(r_q) = \nu_{\odot}
			\begin{cases} 
        			(R_{\odot} / r_q)^{\alpha_{\rm in}}, &  r_q \leqslant r_{\rm break} \\
        			(R_{\odot} / r_b)^{\alpha_{\rm in} - \alpha_{\rm out}} (R_{\odot} / r_q)^{\alpha_{\rm in}}, & r_q > r_{\rm break},
    			\end{cases}   
		\end{equation}
		where $\alpha_{\rm in} = 4.24 \pm 0.08$, $\alpha_{\rm out}=3.21 \pm 0.07$, $r_{\rm break} = 41.3^{+2.5}_{-2.4}$~kpc, $q(\rgce) = 0.86 \pm 0.02$, and $\nu_{\odot}$ the density of
		stars at the Solar radius $R_{\odot}$. Similar to $\nu_{0}$, it is also fix to 1 in our models. In the profile, $R_{\odot}$ is 8.5~kpc. 
\end{itemize}
Fig.~\ref{fig:num_den_mge} presents the MGE results of the number density profiles, which show good agreements with the ones from analytical expressions.

\subsection{Obtaining the PDFs for the Model Parameters}
\label{subsec:likeli}
Given a star $k$ with measured velocity $\boldsymbol{v}_k$ and its error matrix $\boldsymbol{\Delta}_k$ at the position $(R, z)$, the  dynamical probability for tracer $t$ for an assumed Gaussian
velocity distribution is then
\begin{equation}
\label{eq:dyn_prob}
	\begin{aligned}
	P_{{\rm dyn}, k}^t = & \frac{1}{\sqrt{(2 \pi)^3 \det(\boldsymbol{S}_k^t + \boldsymbol{\Delta}_k)}} \times \\
	                            & \exp \left[-\frac{1}{2}(\overline{\boldsymbol{u}}^t - \boldsymbol{v}^t_k)^{\rm T} (\boldsymbol{S}_k^t + \boldsymbol{\Delta}_k)^{-1}(\overline{\boldsymbol{u}}^t - \boldsymbol{v}^t_k)\right], \\
	\end{aligned}
\end{equation}
 where $\boldsymbol{S}_k^t$ is the model predicted velocity dispersion tensor, that is, 
\begin{equation}
\label{eq:vd_mod}
\begin{aligned}
	\boldsymbol{S}_k^t = 
		\begin{bmatrix} 
  			\sigma_{R}^2 & \sigma_{Rz}^2 & 0. \\
  			\sigma_{Rz}^2 & \sigma_{z}^2 & 0. \\
  			0. & 0. & \sigma_{\phi}^2 \\
  		\end{bmatrix}^t_k
		= &
		\begin{bmatrix} 
  			\overline{\uR^2} & \overline{\uR \uz} & 0. \\
  			\overline{\uR v_z} & \overline{\uz^2} & 0. \\
  			0. & 0. & \overline{\up^2} \\
  		\end{bmatrix}^t_k
		- \\
		& \begin{bmatrix} 
  			\overline{\uR}^2 & \overline{\uR}~\overline{\uz} & \overline{\uR}~\overline{\up} \\
			 \overline{\uR}~\overline{\uz} & \overline{\uz}^2 & \overline{\uz}~\overline{\up} \\
			 \overline{\uR}~\overline{\up} & \overline{\uz}~\overline{\up} & \overline{\up}^2 \\
		\end{bmatrix}^t \\
\end{aligned}
\end{equation}
$\overline{\uR}$ and  $\overline{\uz}$ of $\overline{\boldsymbol{u}}^t$ is the mean velocity given by the model. As described at the beginning of this section, in the solution of JAM$_{\rm sph}$, $\overline{\urs} = \overline{\ut} = 0.$~\kms. Hence, $\overline{\uR} = \overline{\uz} = 0.$~\kms after the coordinate transformation. However, as discussed in \citep{capp20}, the Jeans equations constrains the $\overline{\up^2}$ rather than splitting $\overline{\up}$ and $\sigma_{\phi}$ (Eq.~\ref{eq:vphi2}). The value of $\sigma_{\phi}$ is instead defined by the anisotropy parameter $\gamma$ and $\overline{\vrs^2}$ (Eq.~\ref{eq:bg_g}). Therefore, the average values of intrinsic mean $\vp$ measured in Sec.~\ref{subsec:kine} were adopted as $\overline{\up^k}$.

Different types of stars are constrained by the same gravitational potential of the MW, and therefore, solving the Jeans equations for them separately should yield the same mass distribution. To obtain more accurate and self-consistent mass density distribution results, for each model with different $q_h$ form, we will not only marginalized over the parameters for different tracers separately but also perform a combined case that simultaneously fits multiple tracers. Since the absolute number of KG in our sample is much larger than that of BHBs, this would cause the model parameter $\boldsymbol{p}$ to be dominated by KG in the case of simultaneous fit. To avoid this, we introduce a weight function
\begin{equation}
\label{eq:weight}
	w^t = 
	\begin{cases} 
        		1., & t = {\rm KG} \\
		\frac{N_{\rm KG}}{N_{\rm BHB}}, & t = {\rm BHB}
    	\end{cases},
\end{equation}
where $N_{\rm KG}$ and $N_{\rm BHB}$ are the sample sizes of KG and BHB, respectively. Therefore, the logarithm of the likelihood of the data given the parameters $[\boldsymbol{p}, \beta, \gamma]$ can be written as
\begin{equation}
\begin{aligned}
\ln \mathcal{L} = \sum_{t} w^t \left(\sum_{k} \ln P_{{\rm dyn}, k}^t \right)
\end{aligned}
\end{equation}

Here, we summarize the free parameters in our models with two distinct
stellar populations. Under the assumption that the dark matter halo is a spheroidal NFW radial mass density profile
there are potential parameters and two anisotropy ones:
\begin{itemize}
	\item[(1)] $\rho_s$, the scale density in \Msun~pc$^3$;
	\item[(2)] $r_s$, the scale radius in kpc;
	\item[(3)] $q_h$, the flattening of the dark matter halo: spherical case $q_h = 1.$ versus non-spherical ones:
		\begin{itemize}
			\item[(i)] {color{red} $q_h$ is a free parameter, which is a constant along with $\rgce$ (hereafter constant-$q_h$ model)};
			 \item[(ii)]and $q_h$ is a step function along $\rgce$ (hereafter $q_h(\rgce)$ model)
			 	\begin{equation}
				\label{eq:qh_step}
					q_h(\rgce) = 
					\begin{cases}
						q_{h, 1}, & \rgce \leqslant 10.~{\rm kpc} \\
						q_{h, 2}, & 10. < \rgce \leqslant 15.~{\rm kpc} \\
						q_{h, 3}, & 15. < \rgce \leqslant 20.~{\rm kpc} \\
						q_{h, 4}, & 20. < \rgce \leqslant 30.~{\rm kpc} \\
						q_{h, 5}, & \rgce > 30.~{\rm kpc}
					\end{cases}
				\end{equation}
				For the KG sample, as shown in Fig.~\ref{fig:num_den}, stars within 20~kpc dominate. Therefore, the intervals in \rgc
				are smaller in this range to avoid an excessively large sample size in any interval. For BHB stars, this 
				division ensures that the sample sizes in each $\rgce$ interval are as similar as possible.	
		\end{itemize}
	\item[(4)] $\beta$, the axial anisotropy for KG and BHB each;
	\item[(5)] $\gamma$, the tangential anisotropies for KG and BHB each.
\end{itemize}

\section{Results and Discussion}
\label{sec:results}
\begin{figure} [ht!]
\centering
\includegraphics[width=\textwidth]{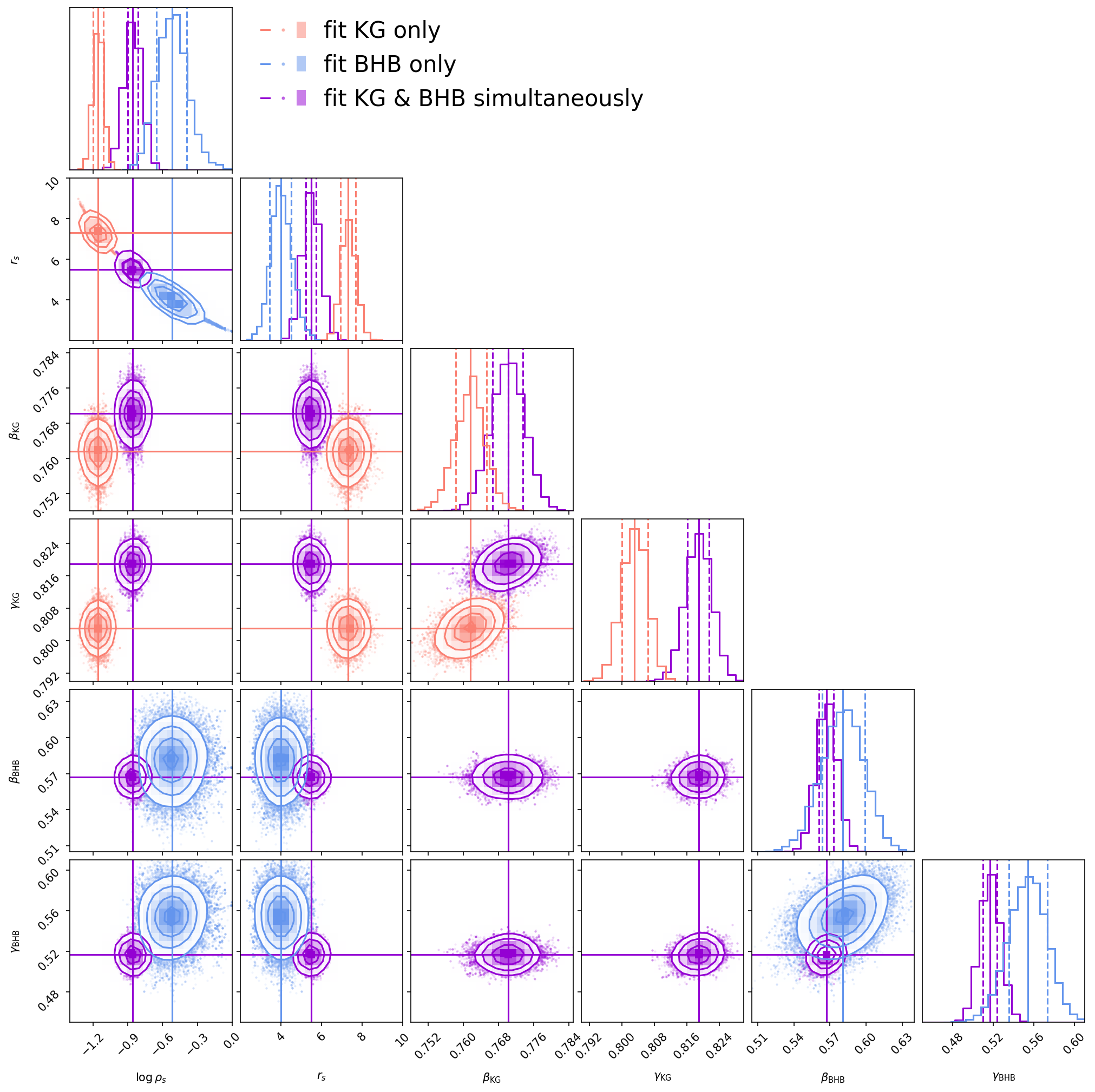}
\caption{One- and two-dimensional projections of the posterior PDFs of the parameters of dark matter density profile with $q_h = 1$.
		The individual fits for KGs and BHBs, and simultaneous fit for both tracers, are all shown in salmon, light blue, and purple, respectively.
		 For all the models, the best fitted values are labeled in thick lines, while the 68\% confidence regions are shown in dashed lines.}	   
\label{fig:pdfs_sphericaldm_bpl}
\end{figure}

\begin{figure} [ht!]
\centering
\includegraphics[width=\textwidth]{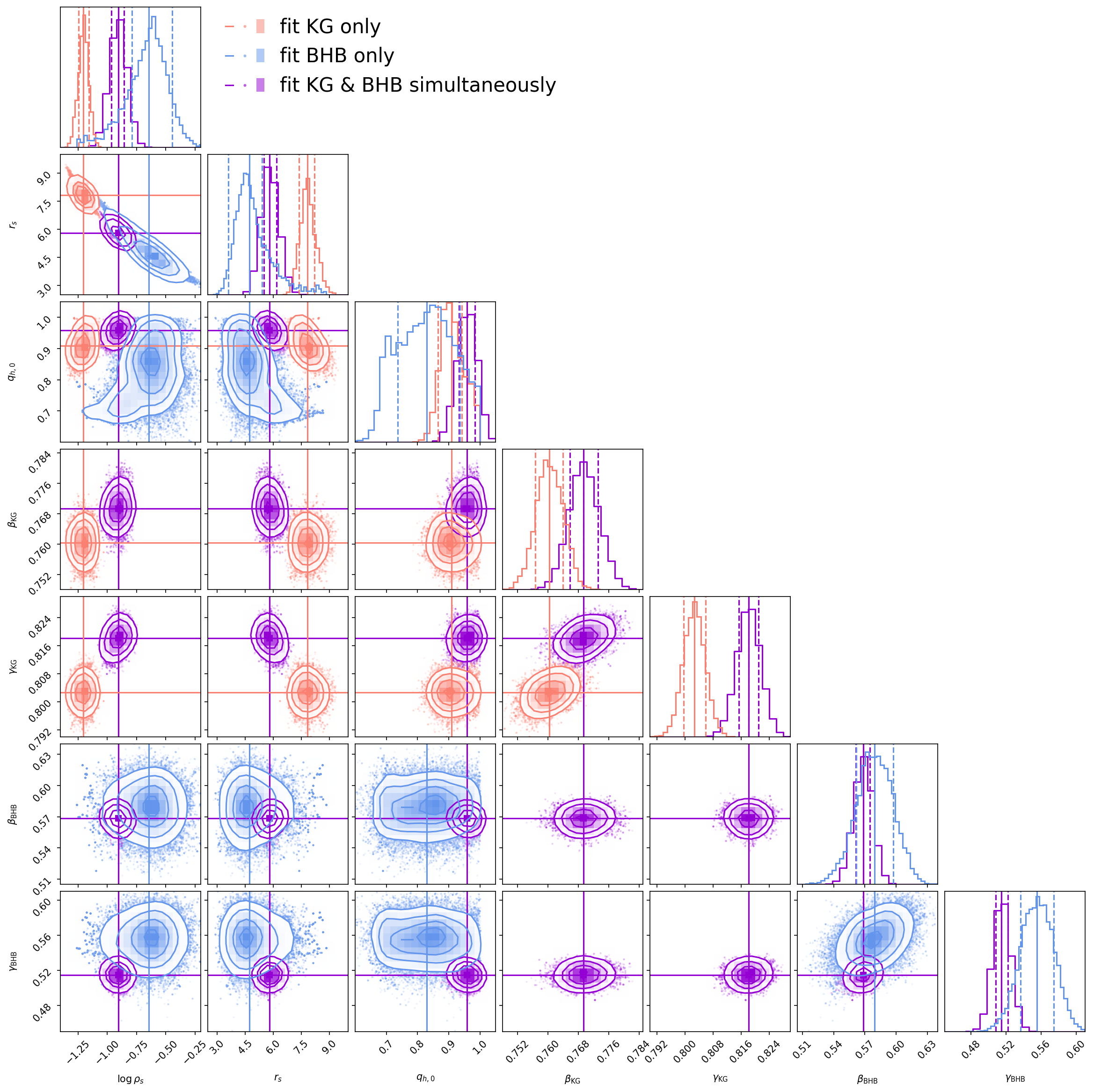}
\caption{Similar to Fig.~\ref{fig:pdfs_sphericaldm_bpl}, but for constant-$q_h$ model.}	   
\label{fig:pdfs_freeq_bpl}
\end{figure}

\begin{figure} [ht!]
\centering
\includegraphics[width=\textwidth]{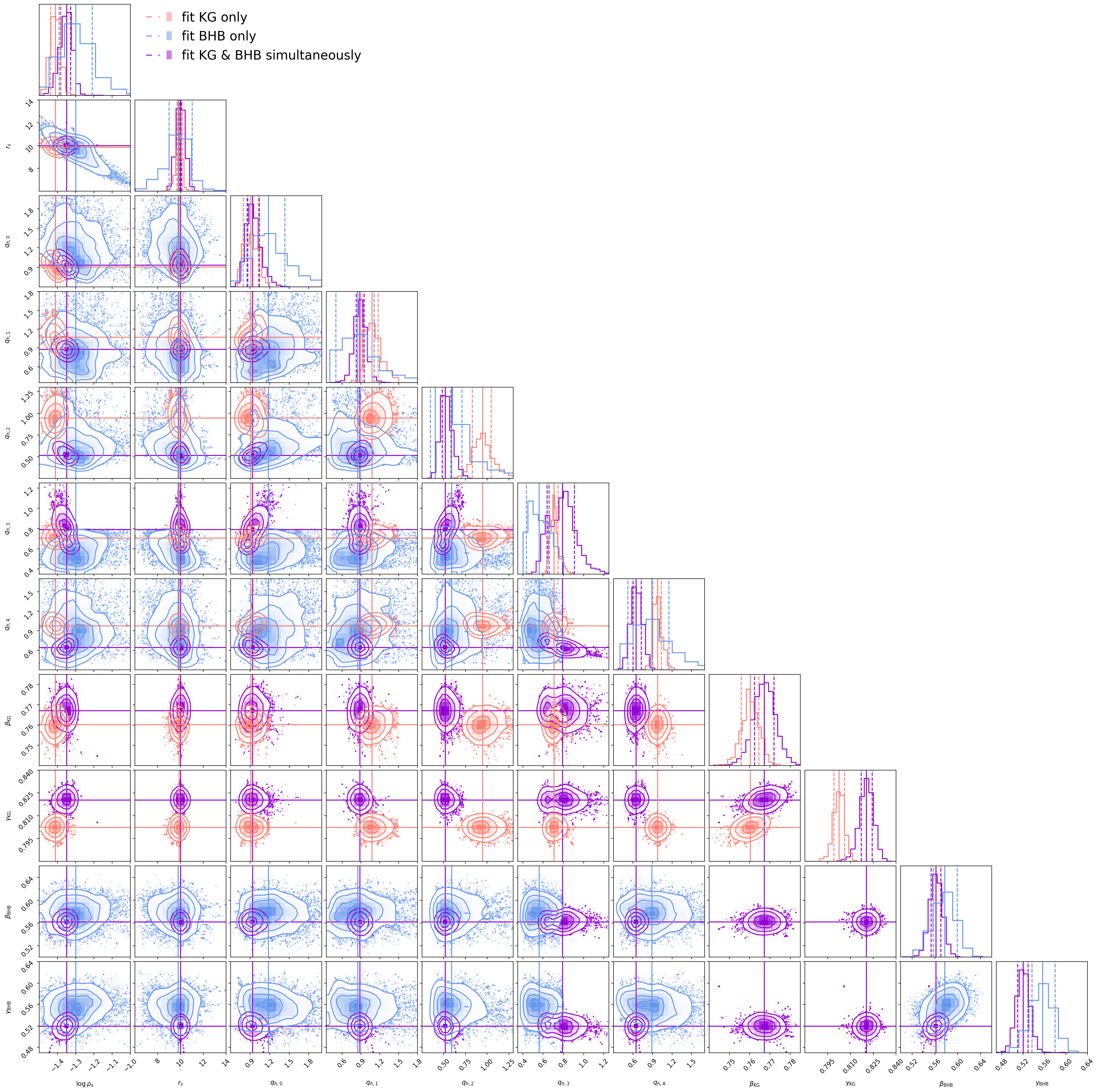}
\caption{Similar to Fig.~\ref{fig:pdfs_sphericaldm_bpl}, but for $q_h(\rgce)$ model.}	   
\label{fig:pdfs_multiq_bpl}
\end{figure}

The above procedure results in PDFs for the parameters of the dark mater density profile and the anisotropy, which are shown in
Figs~\ref{fig:pdfs_sphericaldm_bpl} - \ref{fig:pdfs_multiq_bpl}. The best fitting results and recovered parameters for all dark matter
density models are listed in Tabs.~\ref{tab:best_fit_q1_bpl} - \ref{tab:best_fit_mq_bpl}. The typical fitting errors, estimated from
the last 200 post-burn-in iterations of the MCMC chain where the log probability remained stable with fluctuations below 0.03\% over the final 300 steps, 
are indicated by vertical dashed lines in each posterior distribution plot of the parameters.
For all kinds of models analyzed in this study, the results show the
expected degeneracy between the scale density and the scale radius: a lower $\rho_s$ corresponds to a larger $r_s$. In the constant-$q_h$ model,
there is a degeneracy between $q_h$, $r_s$, and $\rho_s$, though it is less pronounced between the latter two parameters. In contrast, for the $q_h(\rgce)$ model,
the degeneracy mainly occurs between different $q_h$ and $\rho_s$ values, which leads to larger parameter fitting errors in $q_h$.
Moreover,  in this $q_h(\rgce)$ model, the results for the dark matter profile parameters are consistent in 1-$\sigma$ error range between separate fitting and simultaneous fitting for KG and BHB.
 This agreement validates the robustness of the $q_h(\rgce)$ model in describing the dark matter distribution.

The results for the anisotropy parameters $\beta$ and $\gamma$ are not affected by the choice of model. The anisotropy $\beta'$ is $\sim 0.78$ and $\sim 0.57$ for
KG and BHB, respectively. Our results show good consistency with studies of \citet{bird21}, in which the authors studied the anisotropy profiles over all metallicities along
all distances for the two kinds of stars, and concluded that $0.6 < \beta' < 0.9$ for KGs and and a downward shift of $\Delta \beta' \sim 0.1 - 0.3$ for BHBs.

\subsection{Velocity dispersion tensors}
\label{subsec:vd_tensor}
\begin{figure}
	\gridline{\fig{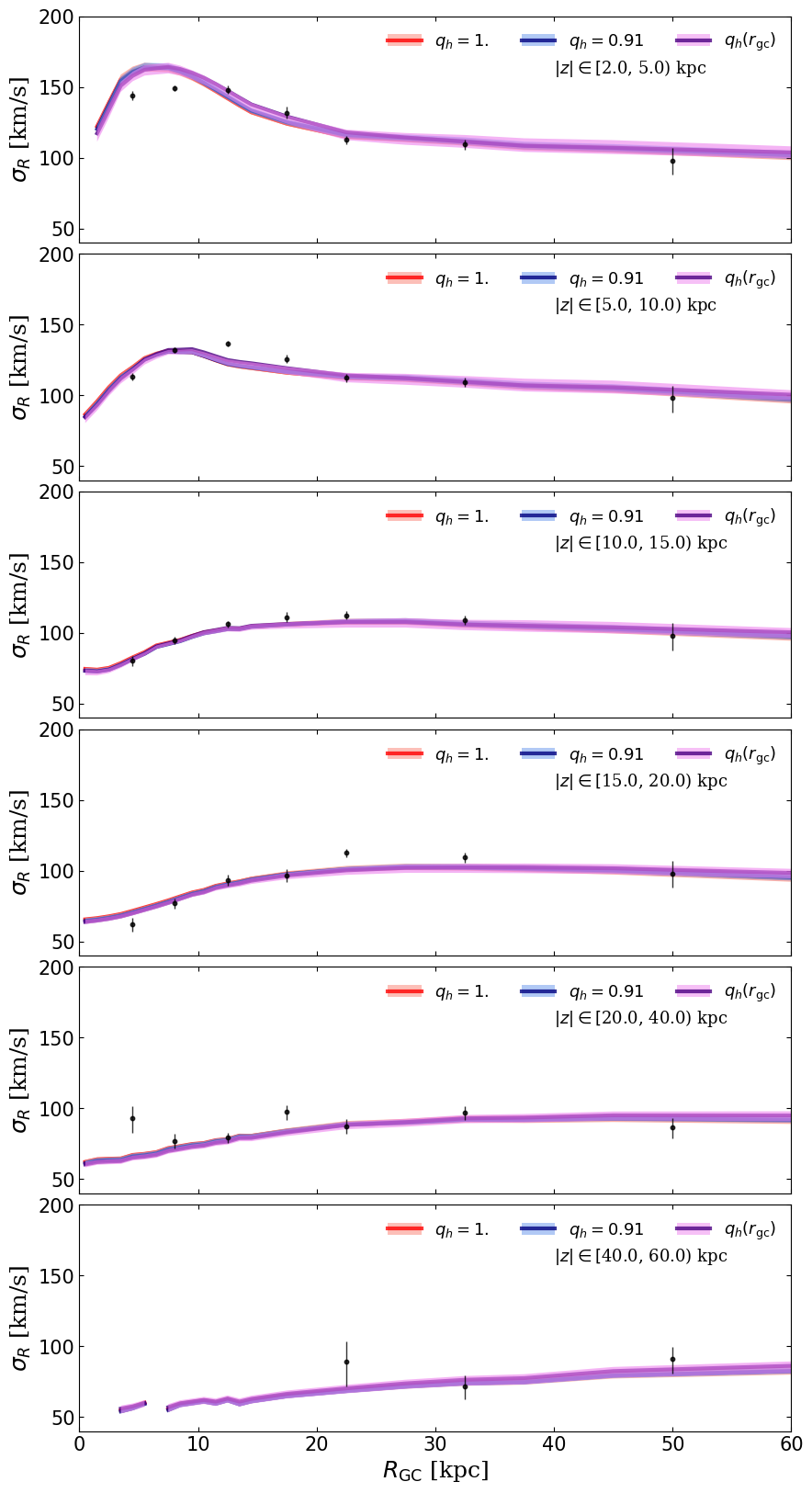}{0.25\textwidth}{}
		     \fig{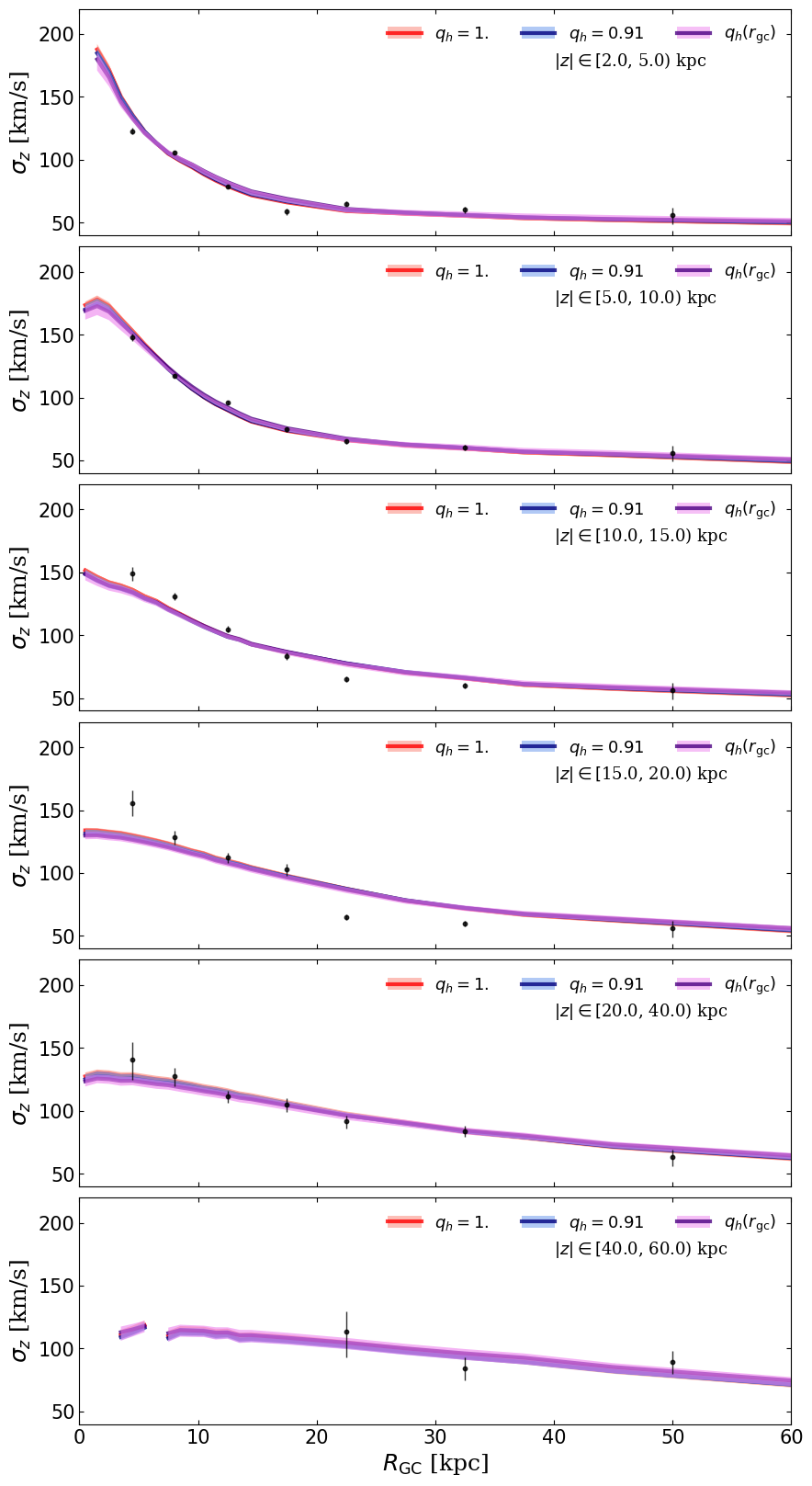}{0.25\textwidth}{}
	            \fig{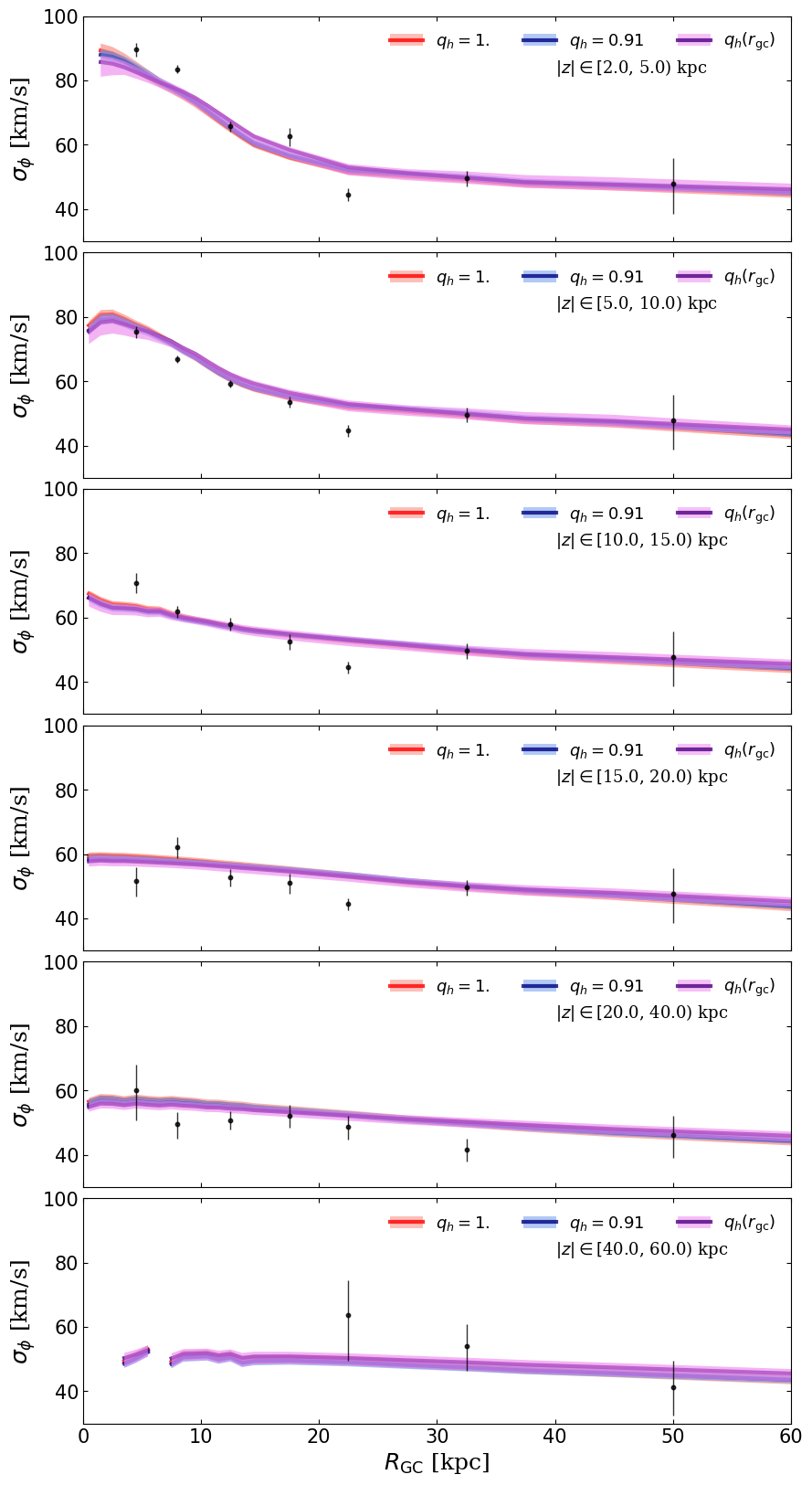}{0.25\textwidth}{}
	            \fig{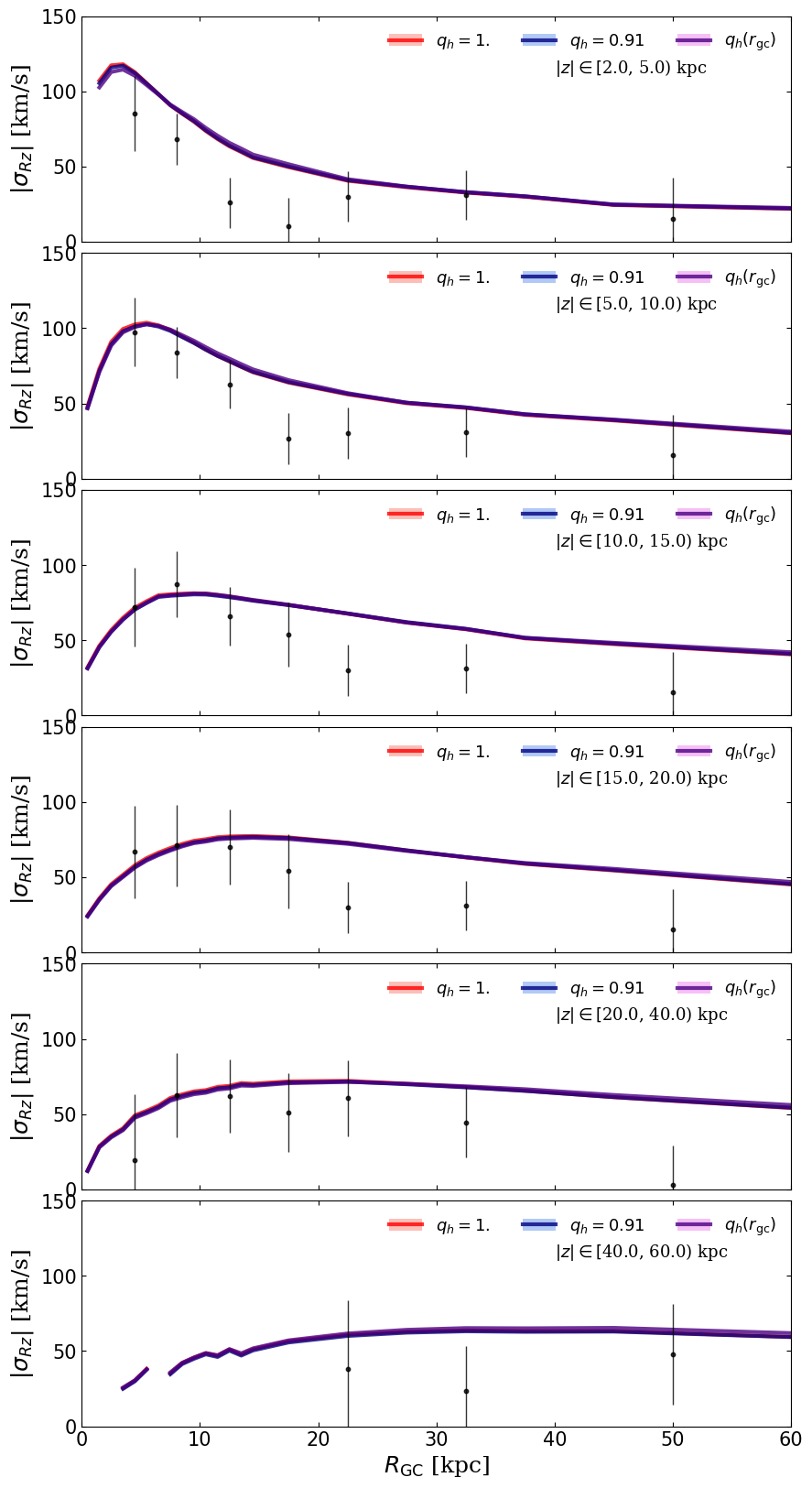}{0.25\textwidth}{}}
    	\caption{Comparisons between components of observed velocity-dispersion tensors and the ones from different model predictions for the KGs,
	             in the case of fitting KG only. From left to right, the plots show $\sR$, $\sz$, $\sph$, $\sigma_{Rz}$. In each panel, the velocity dispersions derived
	             from Eqs.~\ref{eq:v_like}} as well as the corresponded uncertainties, are represented by black dots and the error bars. The red, blue, and purple solid lines,
	             accompanied by shadows, represent  the component values and 1-$\sigma$ error range predicted by
	             the best-fit parameters for different dark matter density profile models, e.g., $q_h = 1$, $q_h=0.91$, and $q_h({\rgce})$, respectively.
\label{fig:vd_compare_kg_sp}
\end{figure}

\begin{figure}
	\gridline{\fig{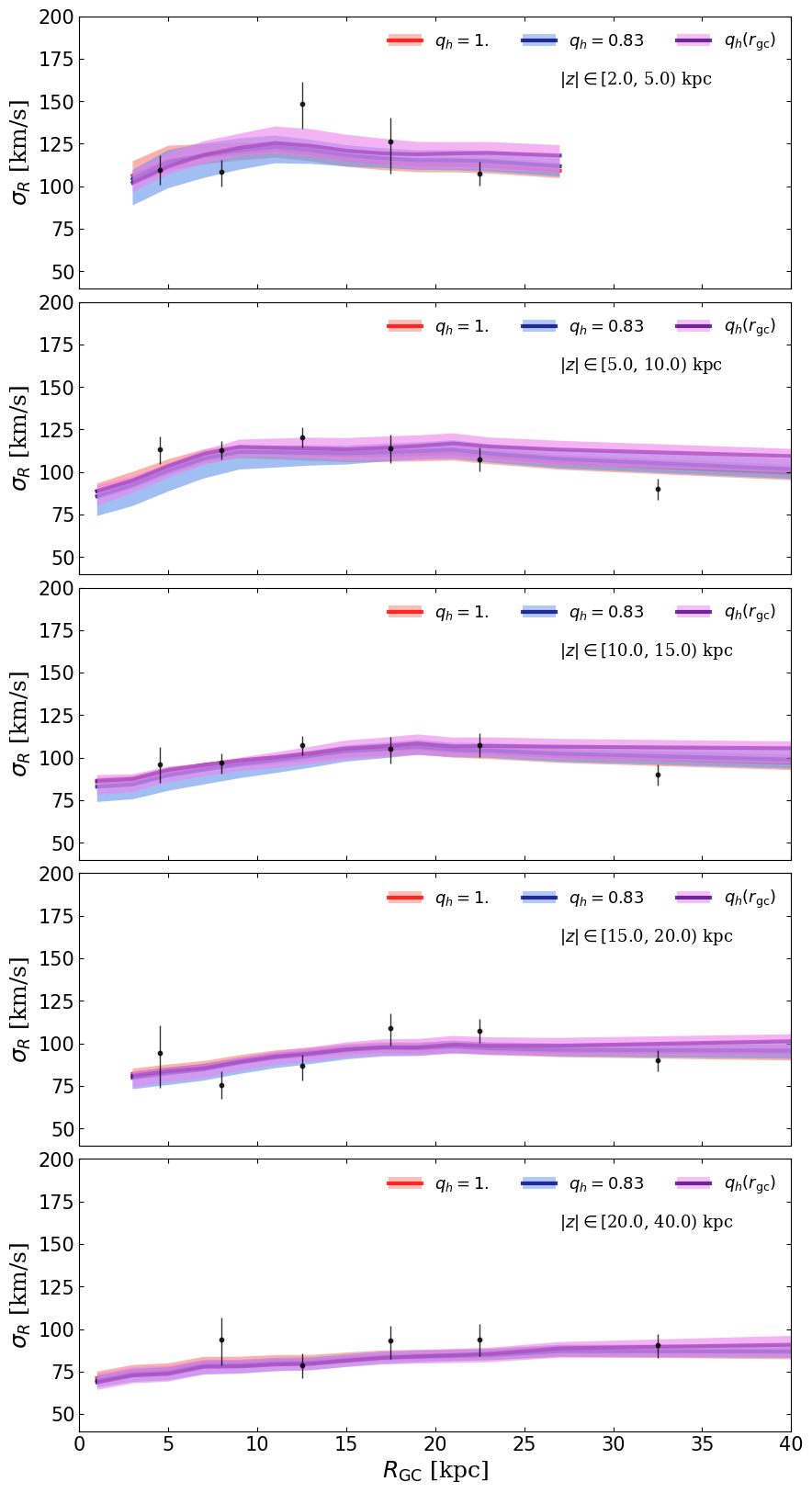}{0.25\textwidth}{}
		     \fig{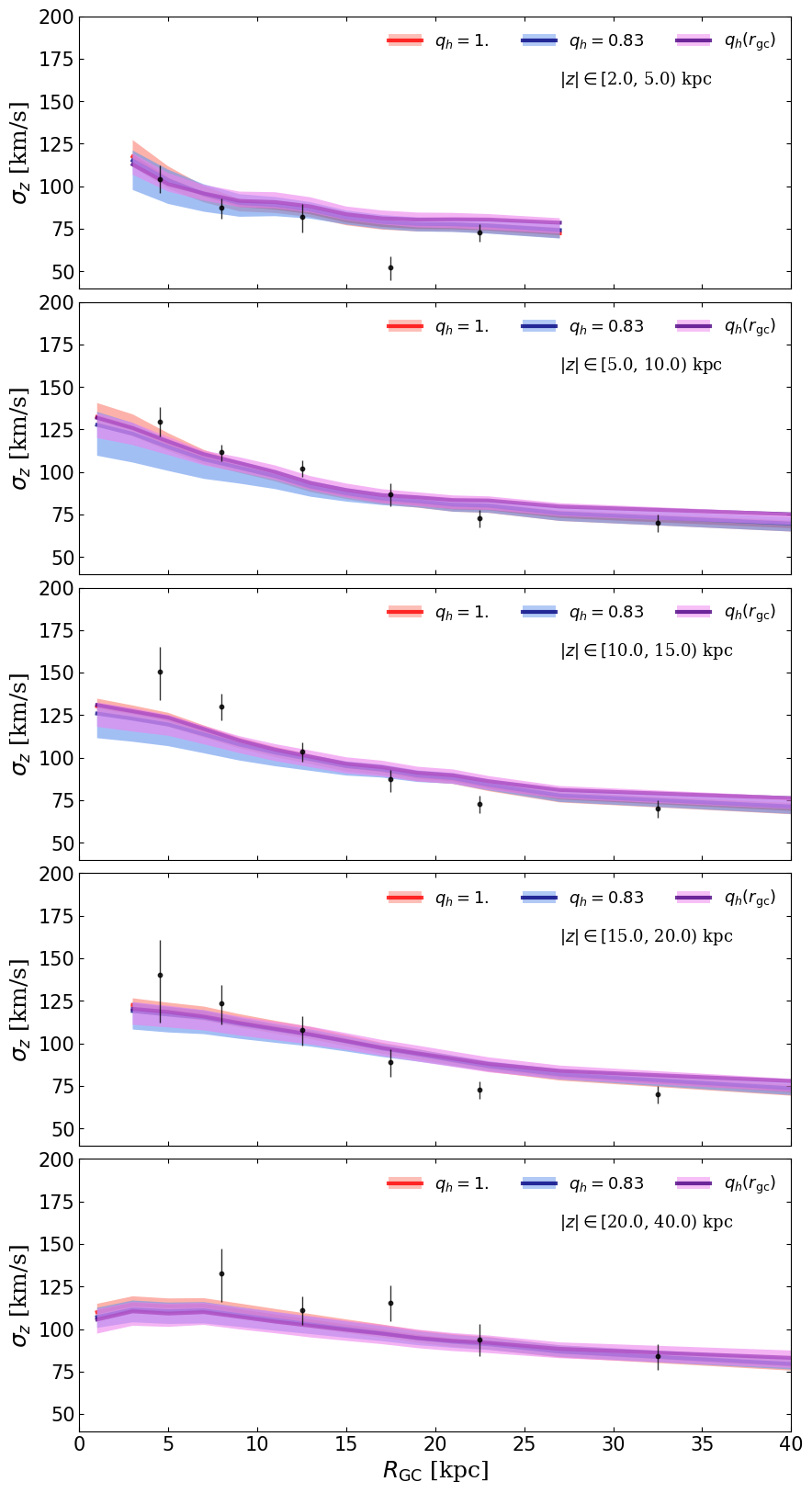}{0.25\textwidth}{}
	            \fig{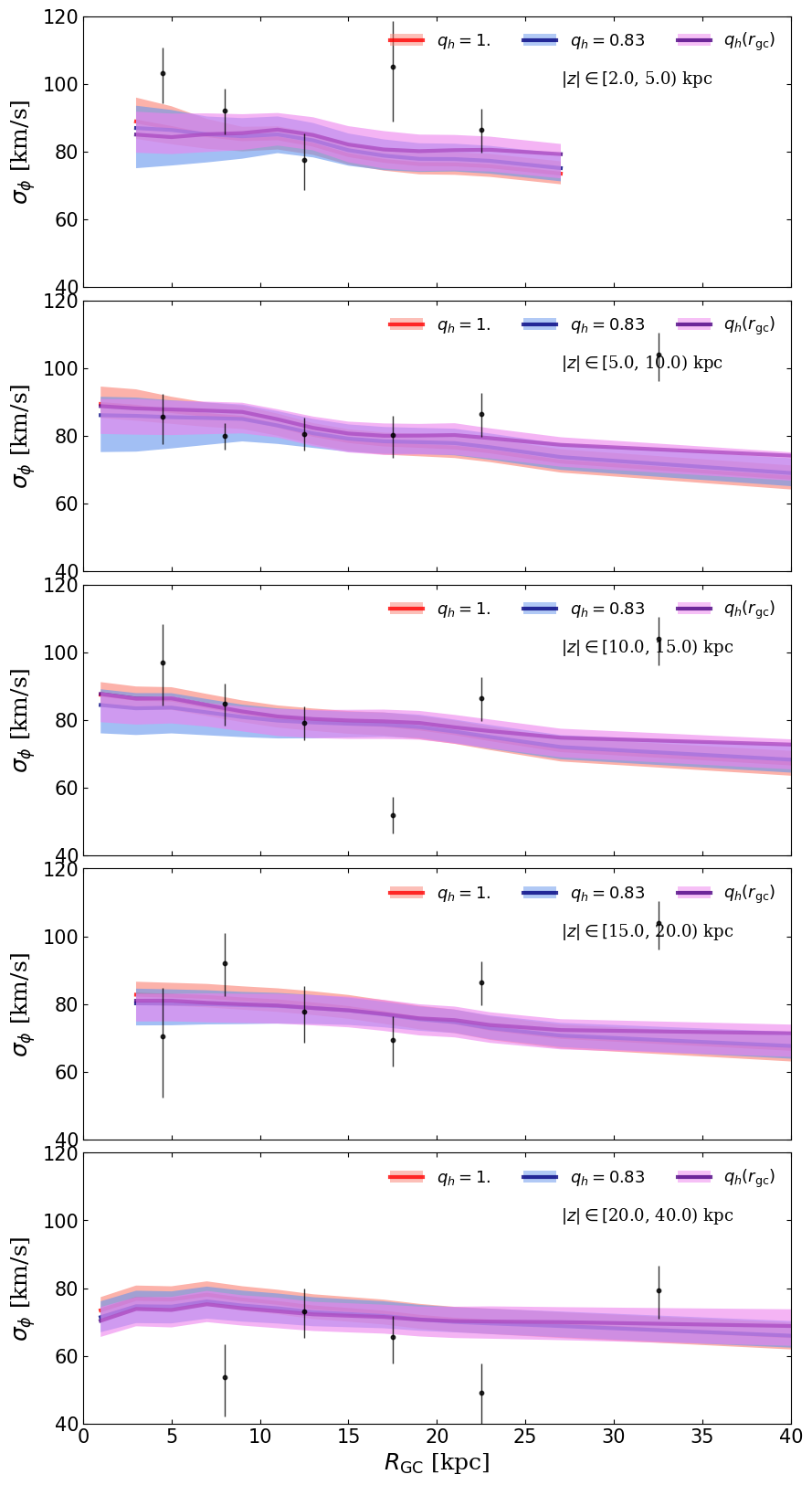}{0.25\textwidth}{}
	            \fig{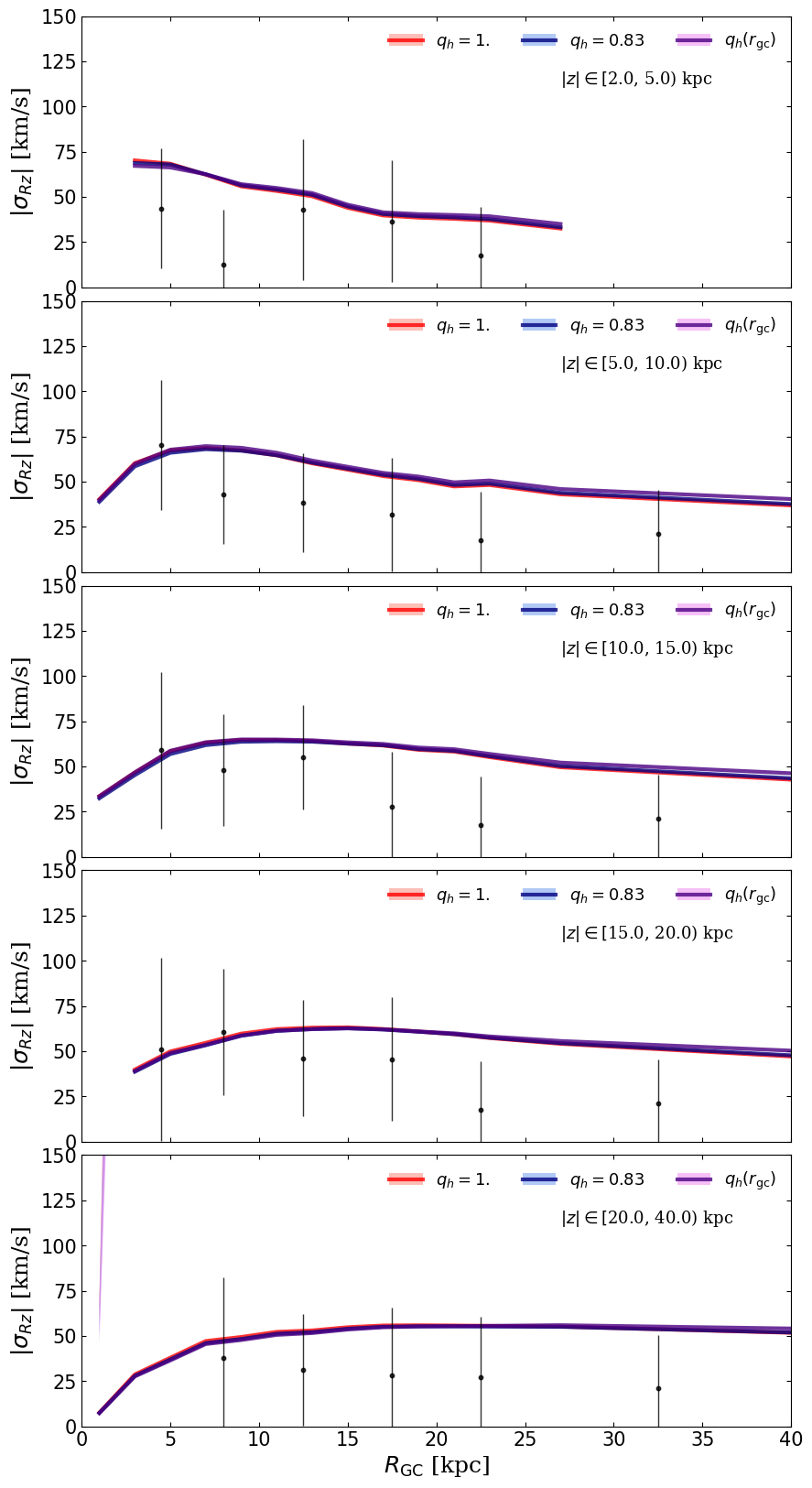}{0.25\textwidth}{}}
    	\caption{Similar to Fig.~\ref{fig:vd_compare_kg_sp}, but for the BHB samples. In this case, if $q_h$ is a constant free parameters along \rgc, $q_h=0.83$.}
\label{fig:vd_compare_bhb_sp}
\end{figure}

\begin{figure}
	\gridline{\fig{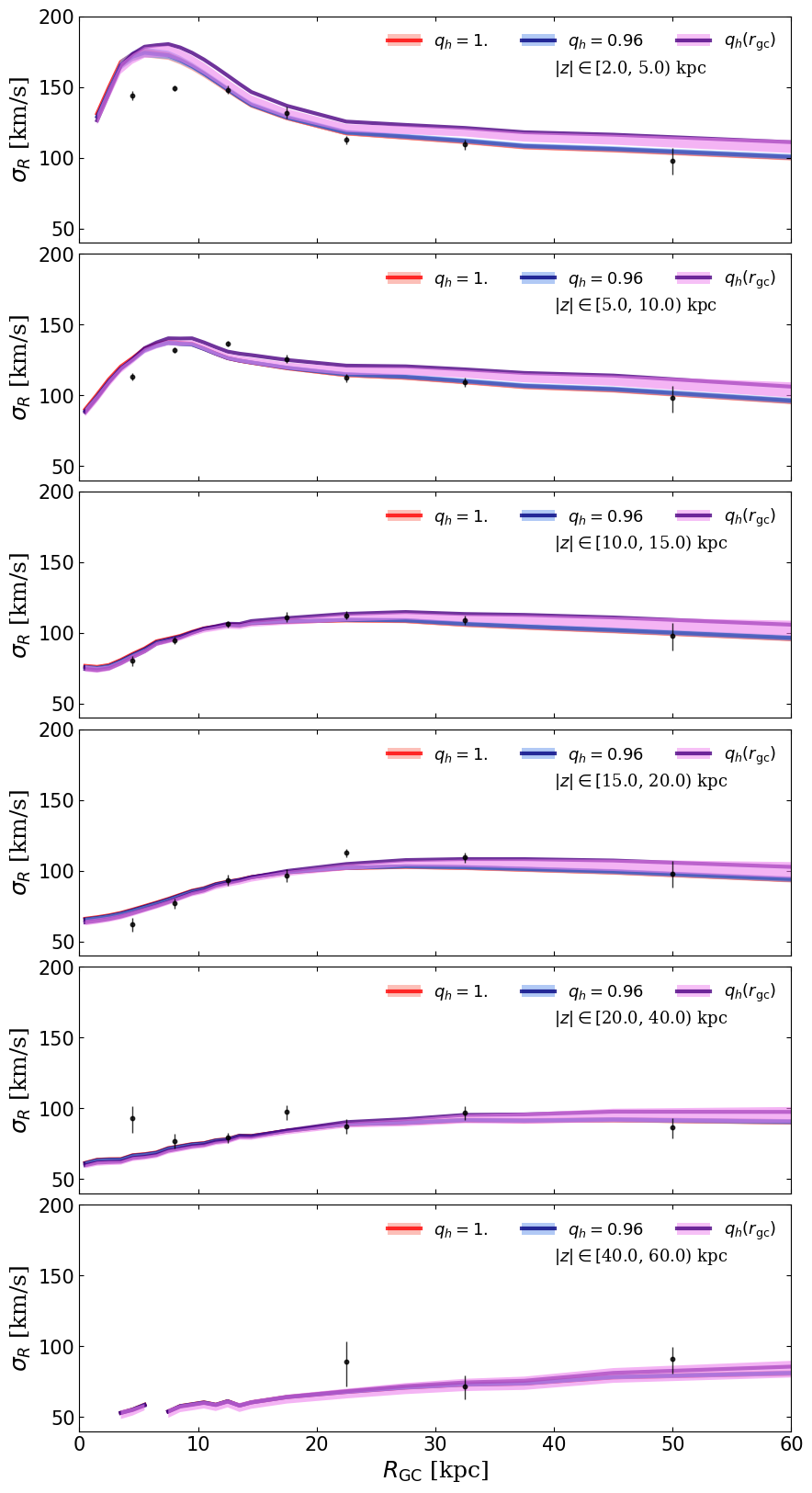}{0.25\textwidth}{}
		     \fig{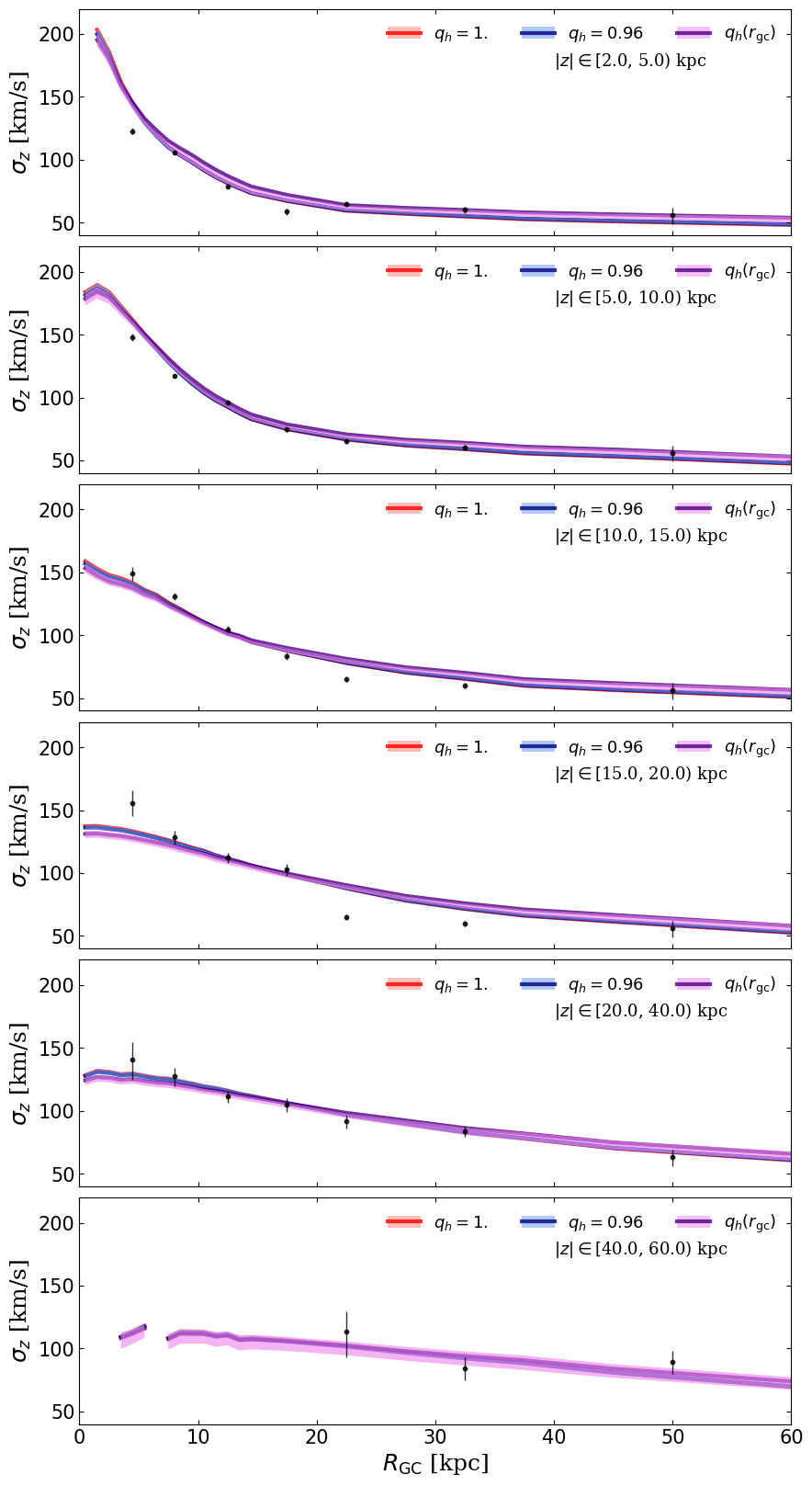}{0.25\textwidth}{}
	            \fig{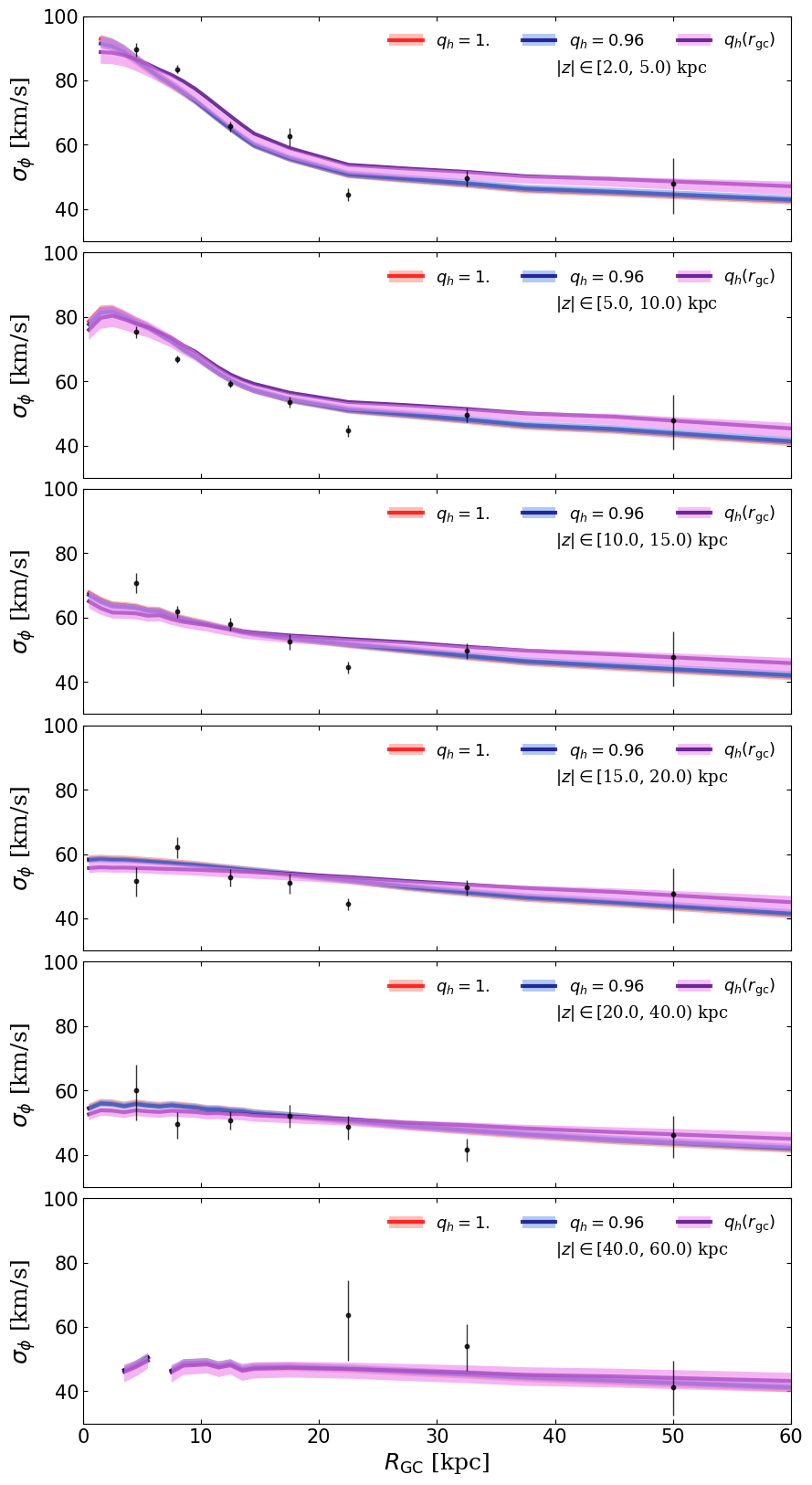}{0.25\textwidth}{}
	            \fig{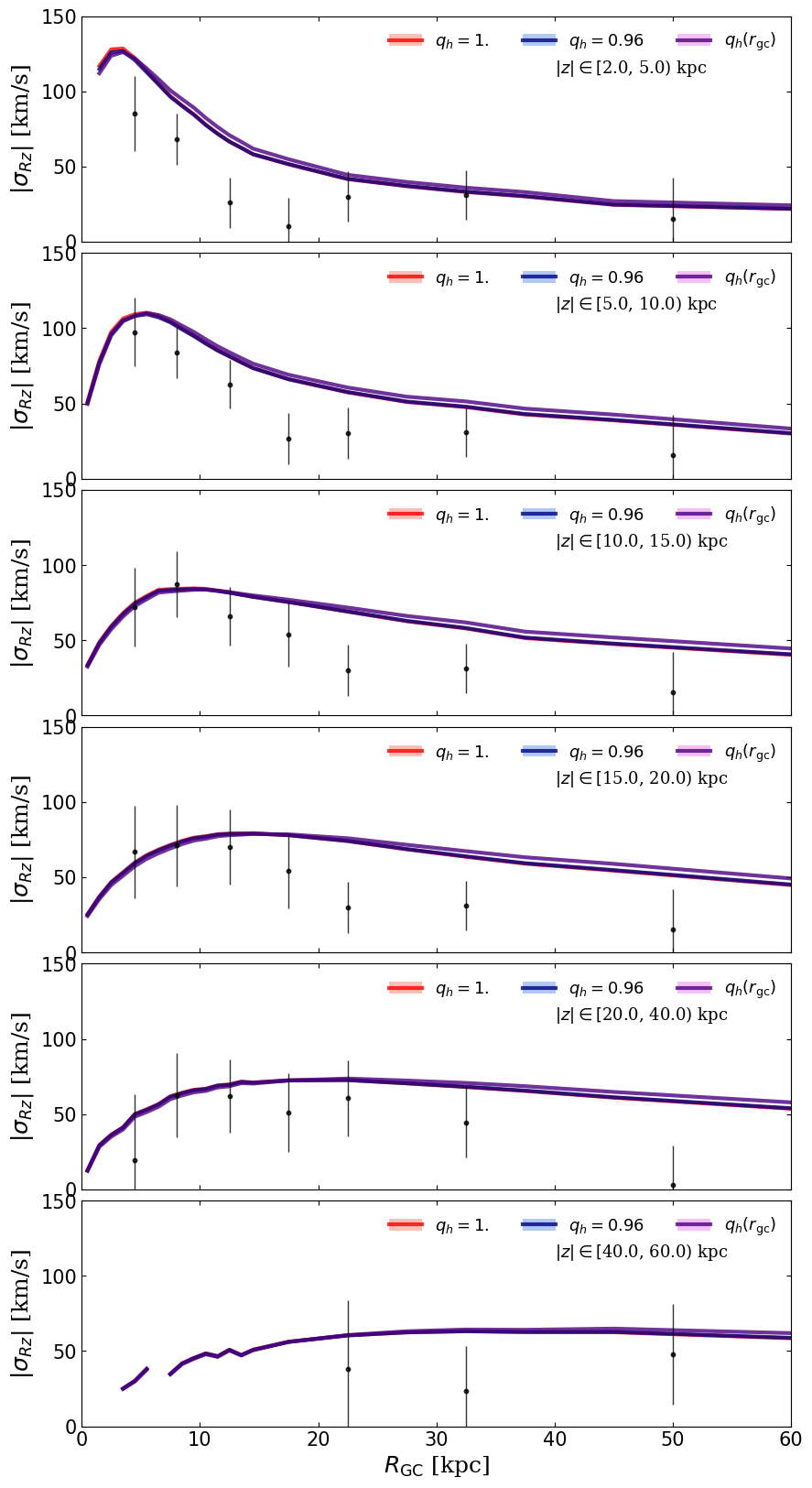}{0.25\textwidth}{}}
    	\caption{Similar to Fig.~\ref{fig:vd_compare_kg_sp}, but for the case of simultaneous fitting.}
\label{fig:vd_compare_kg_mp}
\end{figure}

Figs.~\ref{fig:vd_compare_kg_sp} - \ref{fig:vd_compare_bhb_mp} compare the observed velocity dispersion tensor $\boldsymbol{S}_{\rm obs}$ with the model predict ones
$\boldsymbol{S}_{\rm mod}$. The figures present two sets of results: fitting KG and BHB separately (Figs.~\ref{fig:vd_compare_kg_sp} \& \ref{fig:vd_compare_bhb_sp}),
and simultaneous fitting (Figs.~\ref{fig:vd_compare_kg_mp} \& \ref{fig:vd_compare_bhb_mp}). Components of
$\boldsymbol{S}_{\rm obs}$ have been corrected for the velocity errors of the individual stars within each $R-|z|$ cell,
as discussed in Sec.~\ref{subsec:kine}. 

In these plots, we also summarize model values $\boldsymbol{S}_{\rm mod}$ predicted by different $\rho_{\rm DM}$ models for the most likely parameters,
which are shown in thick lines. It can be seen that the model predict trends show good agreements with the observed ones.
Uncertainties in the model predict values are mainly
caused by (1) observed errors from kinematics; (2) uncertainties of the fitting process; (3) errors in parameters of number density profiles. Error matrix from observed kinematics, $\boldsymbol{\Delta}$, derived using the method described in Sec.~\ref{subsec:kine}, has been accounted in the calculation the dynamical probability of each star (Eq.~\ref{eq:dyn_prob}). These errors will propagate into the final fitting results, and reduce the weight of stars with larger errors in the likelihood calculation.


To estimate the effect from the errors of the number density profiles, we systematically vary all parameters in $\nu_{\star}$ within the uncertainties listed in Sec.~\ref{subsec:nd} and repeat the fitting process. Our analysis shows that the final results remain unsensitive by these small variations ($\sim 2\%$). In particular,
the relative uncertainties are typically below 0.1\%, which is can be neglected.
Among the models, $q_h(\rgce)$ model yields the largest fitting uncertainties. However, even in this case, it is still smaller than the errors of of $\boldsymbol{S}_{\rm obs}$. 
In the same set of $\rho_{\rm DM}$ model, the largest fitting errors were obtained in the case of fitting the BHB seperately, which is due to the smaller sample size of the BHB stars. 

Overall, it can be seen that the moments of velocity dispersion tensor predicted by the Jeans model show good agreement with observations within 1-$\sigma$ uncertainties for all components except $\cRz$.
This discrepancy arises because JAM$_{\rm sph}$, when solving the axisymmetric Jeans equation, assumes that the velocity ellipsoid is aligned with the spherical coordinate system \citep{capp20}. 
This divergence reflects the complexity of real galactic dynamics: observational data indicate that the velocity ellipsoid of the MW's halo cannot be globally described by either spherically aligned or cylindrically aligned coordinate systems. This suggests the inherent limitations of classical Jeans analysis in capturing non-axisymmetric kinematic structures.

We selected a sample from one $R-|z|$ cell, e.g., $R \in [10., 15.]$~kpc and $|z| \in [15., 20.]$~kpc, using the simultaneously fitting results as an example shown in
Fig.~\ref{fig:vd_examples}. The histograms along the diagonal represent the velocity distributions of the sample stars in different velocity components, while the scatter plots off-diagonal show the distributions in 2D velocity spaces. The predicted velocity distributions at each star's location, as predicted by the models, are shown as solid lines of different colors. All three models from this work clearly reproduce the observed velocity distributions well.

The results from separate and simultaneous fits of KG and BHB are very close, with differences smaller than the error of observed
indicating a good consistency in the current models. Additionally, for the same tracer, the differences in given by different models are also smaller than the error of $\boldsymbol{S}_{\rm obs}$.

\subsection{Dark matter distribution}
\label{subsec:dm_profile}
\subsubsection{Enclosed mass profile}
\label{subsubsec:en_mass}
The $q_h(\rgce)$ model stands out as the most reliable, as it yields consistent $\rho_s$ and $r_s$ values across KG, BHB, and simultaneous fitting, implying a varying 
$q_h(\rgce)$ along \rgc. In contrast, the spherical ($q_h$ = 1) and  constant-$q_h$ model show a lack of alignment in $\rho_s$ and $r_s$ within 1-$\sigma$ error range
for KG and BHB individually. Furthermore, the $r_s$ value derived from BHB in these models is less than 5~kpc, which is inconsistent with the lower distance limit of BHB stars. While the 
$q_h(\rgce)$ model exhibits larger fitting errors due to the smaller sample size in the outer halo. 


On the basis of the dark matter density profiles, combing with baryonic matter density described in Sec.~\ref{subsec:gp}, we can derive the total enclosed mass at any $\rgce$ with
\begin{equation}
\label{eq:en_mass}
	M_{\rm tot} (< \rgce) = 4 \pi \int_{0}^{\rgce} \int_{0}^{\sqrt{\rgce^2 - R^2}} \rho_{\rm tot} (R, z) R dz dR
\end{equation}
The mass results are shown in  Fig.~\ref{fig:rho_mass}. The red and blue thick lines represent the results calculated by the most likely parameter results from separately fitting
KG and BHB, respectively, while the purple lines correspond to the simultaneous fits. The bands of lines show mass results from 1-$\sigma$ sampling of the PDF of $\rho_{\rm DM}$.

According to the definition of NFW, $\rho_s = \rho_{\rm cirt} \delta_c$, where $\rho_{\rm cirt} = 3 H_0^2 / (8 \pi G)$ \citep[$H_0 = 73.01\pm 0.92$~${\rm km~s^{-1}~Mpc^{-1}}$, ][]{mura23}
is the background density at the time of the halo formation and $\delta_c$ is a characteristic overdensity. Therefore, we can estimate the viral radius $r_{200}$ through the definition that it is the radius with mean overdensity 200, and hence the dark matter viral mass. The results are also listed in Tabs.~\ref{tab:best_fit_q1_bpl} - \ref{tab:best_fit_mq_bpl}. All three models present similar characteristics: the errors from fitting BHB separately are larger, primarily due to the smaller BHB sample size. Within the 1-$\sigma$ error range, KG and BHB produce nearly consistent results, and the simultaneous fits show good agreements with the separate fits. 

%

Since our sample is primarily concentrated within 60 kpc, and to facilitate comparison with literature studies, we also calculated the total mass within 40 kpc and 60 kpc using the simultaneous fitting results. $M_{\rm tot}^{q_h=1} (< 40~{\rm kpc}) = 0.416^{+0.012}_{0.005} \times 10^{12}$~\Msun \,\,and
$M_{\rm tot}^{q_h=1} (< 60~{\rm kpc}) = 0.508^{+0.019}_{0.011} \times 10^{12}$~\Msun \,\,for $q_h=1$ model, 
$M_{\rm tot}^{{\rm constant}-q_h} (< 40~{\rm kpc}) = 0.415^{+0.011}_{0.009} \times 10^{12}$~\Msun \,\, and 
$M_{\rm tot}^{{\rm constant}-q_h} (< 60~{\rm kpc}) = 0.509^{+0.017}_{0.013} \times 10^{12}$~\Msun \,\, for constant-$q_h$ model, and
$M_{\rm tot}^{q_h({\rgce})} (< 40~{\rm kpc}) = 0.428^{+0.020}_{0.031} \times 10^{12}$~\Msun \,\, and 
$M_{\rm tot}^{q_h({\rgce})} (< 60~{\rm kpc}) = 0.533^{+0.061}_{0.054} \times 10^{12}$~\Msun \,\, for $q_h(\rgce)$ model.

While posterior uncertainties grow at large $\rgce$ due to sparse tracers, all models yield statistically consistent $M_{\rm tot}(< \rgce)$ within 1-$\sigma$. Within the uncertainties, these are consistent with 
$M_{\rm tot}(< 39.5~{\rm kpc}) = 0.42^{+0.07}_{-0.06} \times 10^{12}$~\Msun \,\, from \citet[][hereafter W19]{watkins19} by estimating from halo globular clusters, but slightly higher than
$M_{\rm tot}(< 60~{\rm kpc}) = 0.4\pm0.07 \times 10^{12}$~\Msun from X08 by using BHB stars from SDSS DR6.

In addition to the results from W19 and X08, we also include the calculations from B22, which used both KG and BHB while solving the spherical Jeans equations.
W19, X08, and B22 all assume a spherical dark matter halo. Moreover, B22 also assumes a spherical number density profile. It can be seen that under these assumptions, the mass estimates for KG and BHB show significant divergence at $\rgce > 20$~kpc. However, our results from solving the axisymmetric Jeans equation show that KG and BHB display good consistency, regardless of $\rho_{\rm DM}$ model. Combined with the fitting results of $\rho_s$ and $r_s$ of different models, this suggests that the Galactic dark matter halo is more likely to be spheroidal in shape.

Compare with the viral mass estimated in B22, our results show good consistency between the tracers of KGs and BHBs. Moreover, in our 
analysis, $M_{200}$ from our models by simultaneous fitting are $M_{200}^{q_h=1}= 0.728^{+0.046}_{-0.034} \times 10^{12}$~\Msun,
$M_{200}^{{\rm constant}-q_h} = 0.736^{+0.046}_{-0.037} \times 10^{12}$~\Msun, and
$M_{200}^{q_h({\rgce})} = 0.820^{+0.210}_{-0.816} \times 10^{12}$~\Msun \,\, for $q_h=1$, constant-$q_h$, and $q_h(\rgce)$ models, respectively. The values are lower than
$M_{200} = 1.54^{+0.75}_{-0.44} \times 10^{12}$~\Msun \,\, of W19 and 
$M_{200} = 1.00^{+0.30}_{-0.20} \times 10^{12}$~\Msun \,\, of X08, but in good agreement with $M_{200} = 0.805 \pm 0.115 \times 10^{12}$~\Msun \,\ of \citet{zhou23} by studying
the rotation curve of luminous red giant branch (LRGB) stars.  More details of our results can be found in Tabs.~\ref{tab:best_fit_q1_bpl} - \ref{tab:best_fit_mq_bpl}.

The comparison of enclosed mass and circular velocity among different $\rho_{\rm DM}$ models from simultaneous fitting results are shown in Fig.~\ref{fig:model_compare}, as well as the
contributions from baryonic matters. There is no significant difference in the $M_{200}$ estimates derived from different $\rho_{\rm DM}$ models.



In comparison to most previous studies, we selected more observationally consistent axisymmetric models for both the number density and dark matter density profiles.
By fitting the velocity-dispersion tensors for different tracers, we obtained consistent results for the dark matter density $\rho_{\rm DM}$ and enclosed mass $M(< \rgce)$, 
indicating that these tracers are 
under the same gravitational potential. Moreover, the results from simultaneously fitting different tracers are also agree with these findings, demonstrating the consistency
and reliability of our current methodology.

Finally, we used the fitted dark matter density profiles to calculate the dark matter density in the solar neighborhood, despite the lack of data close to the Galactic disk.
The results, which are also listed in Tabs.~\ref{tab:best_fit_q1_bpl} - \ref{tab:best_fit_mq_bpl}, present little differences in 1-$\sigma$ error range, e.g., $\rho_{\rm DM, \odot} \sim 0.0135 - 0.0171$~\Msun~pc$^{-3}$, which agree well with $\rho_{\rm DM, \odot} = 0.0151^{+0.0050}_{-0.0051}$~\Msun~pc$^{-3}$ from \citet{guo22} by using G/K-dwarfs from LAMOST DR5.

\subsubsection{Shape of the dark matter halo}
\label{subsubsec:shape}
We also explored the shape of the dark matter halo. In our study, the current results indicate that the flattening of the halo does not significantly affect the $\rho_{\rm DM}$ profile and virial mass. That is, given a number density distribution, different models produce consistent density profiles and mass estimates within the 1-$\sigma$ range. However, when $q_h$ is allowed to vary as a free parameter or a parameterized function, we found that the dark matter halo deviates from a spherical shape. 
our results suggest that the fitted values of $\rho_s$ and $r_s$ converge to consistent results across different tracers only in the $q_h(\rgce)$ model.



Fig.~\ref{fig:q_compare} compares the halo flattening parameter $q_h$ inferred from different tracers and models. Our analysis reveals two key features: 
\begin{itemize}
	\item[(1)] a non-monotonic trend: $q_h$ decreases within $\rgce < 20$~kpc (inner halo), followed by a gradual return to sphericity at larger radii; 
	\item[(2)] the average $q_h$ in the inner and outer regions differ systematically. For KGs, we find $<q_h>_{\rm in, KG} = 0.98 \pm 0.17$ \& $<q_h>_{\rm out, KG} = 0.84 \pm 0.13$; 
		    for BHBs $<q_h>_{\rm in, BHB} = 0.87 \pm 0.40$ \& $<q_h>_{\rm out, BHB} = 0.72 \pm 0.39$; and for our simultaneous multi‑population fit $<q_h>_{\rm in, MP} = 0.78 \pm 0.19$
		    \& $<q_h>_{\rm out, MP} = 0.72 \pm 0.17$.
\end{itemize}

This behavior mirrors recent cosmological hydrodynamical simulations that incorporate baryonic physics but exhibits notable differences in radial scaling. For instance,
Analyzing the Illustris simulation suite \citep{gene14, voge14a, voge14b, sija15}, \citet{chua19} assessed how baryonic physics influence DM halo shapes and showed that MW-analogues exhibit systematically rounder halo‐shape ($q_h = 0.88 \pm 0.10$) than the ones of DM-only simulations ($q_h = 0.67 \pm 0.14$).
Likewise, \citet{shao21}, using MW-mass haloes from EAGLE cosmological hydrodynamics simulation \citep{crain15, scha15}, \citet{shao21}, found that the dark matter halo is not homologous and its flattening vary with radius (black thick line in Fig.~\ref{fig:q_compare}) . The inner parts of the halo are rounder than the outer parts, because the inner haloes in simulations that include baryons are typically rounder than in DM-only simulations. 

In DM–only simulations, DM haloes are generally triaxial, with most exhibiting a slightly prolate shape $(a > b \simeq c)$, where $a$, $b$, and $c$ are the major, intermediate, and minor axes, respectively, is the major as fully confirmed by numerous early studies \citep[e.g., ][]{fren88}. Subsequent cosmological simulations have shown that the inclusion of baryons can significantly influence halo shape: the presence of a central galaxy tends to make haloes rounder compared to their DM-only counterparts \citep{bail05, vell15, chua19, prad19}. Many studies investigating the shape of the Galactic DM halo utilize the tidal stream of the Sagittarius dwarf, which traces the Galactic potential within $\sim 100$~kpc. These works often suggest a highly flattened halo, oriented perpendicular to the Milky Way disc \citep{hemi04, john05, law10, deg13}. The best-fitting model by \citet{law10} describes an oblate halo with axis ratios of $<c/a> = 0.72$ and $<b/a> = 0.99$—flatter than a typical $\Lambda$CDM halo \citep{haya07}. However, studies such as \citet{vera13} and \cite{gome15} have highlighted that the Large Magellanic Cloud (LMC), which is thought to be very massive, can induce significant dynamical perturbations to the orbits of the Sagittarius stream as well as other stellar streams \citep[e.g., the Tucana III stream;][]{erka18}, thereby complicating the modelling of the Galactic halo’s shape. 

In our result, at larger distances, the haloes become systematically more flattened. With the dominant effect being the potential of the baryons, which is very important for $ \rgce / r_{200} < 0.2$, consistent with our observed $q_h$ minimum at  $\rgce / r_{200} \sim 0.1$. This may be because, at this location, the baryonic mass/total mass ratio drops below 23\% (see the upper panel of Fig.~\ref{fig:model_compare}), which can no longer dominate the overall gravitational potential.

\section{Conclusions}
\label{sec:conclusion}
We have analyzed the KG from LAMOST SDSS/SEGUE and BHB form SDSS to determine the density profile, the shape and the viral mass of the MW dark matter halo.
Combining with precise proper motion measurements from Gaia DR3, we investigated the distribution of velocity ellipsoids for KGs and BHBs in the $(R, |z|)$ space. The observed velocity ellipsoids projected onto the $(\vR, \vz)$ plane exhibit nearly spherical alignment, while the velocity ellipses in the $(\vR, \vp)$ and $(\vz, \vp)$ planes align with the cylindrical coordinate system.

Based on these observations, we solved the axisymmetric Jeans equations using the MGE and JAM$_{\rm sph}$ methods, adopting a number density profiles that have been corrected
for selection effects, as determined by previous literature studies, and assuming parameterized $\rho_{\rm DM}$ with different flattening models, to fit the velocity distribution at each sample
star's spatial location. Then an MCMC technique is employed to recover the PDFs of the parameters of $\rho_{\rm DM}$.

For each $\rho_{\rm DM}$ model, we performed two types of fits: individual fits for KGs and BHBs and the simultaneous fit both, to examine whether an axisymmetric assumption could reduce the discrepancies in the results produced by different tracers. Our results show good consistency over different tracers, dark matter flattening models, and fitting methods, yielding consistent density profiles, $r_{200}$, and $M_{200}$ within a 1-$\sigma$ confidence region.
Moreover, the model with a varying halo flattening parameter with respect to \rgc ~showed the best consistency, implying that the shape of the MW dark matter halo may change with \rgc. This result aligns with simulated results from \citet{shao21}. In this model, the derived $r_{200}$ is $188^{+15}_{-15}$~kpc, corresponding to an $M_{200}$ of $0.820^{+0.210}_{-0.816}$~\Msun. Finally, using the fitted parameters, we  estimated the local dark matter density of $\rho_{\rm DM, \odot} \sim 0.0135 - 0.0171$~\Msun~pc$^{-3}$.

\begin{acknowledgments}
This work is supported by National Key Research and Development Program of China No. 2019YFA0405504,
National Natural Science Foundation of China (NSFC) under grants No. 11988101, and 11873052, and China Manned Space Project with No. CMS-CSST-2021-B03.
Lan Z., X-X.X., Ling Z. and S.S. acknowledge the support from CAS Project for Young Scientists in Basic Research Grant No. YSBR-062, and the hospitality of the International
Center of the Supernovae (ICESUN).
S.S. acknowledge the support from NSFC under grants No. 12273053.
C.Y. acknowledge the support from NSFC under grants No. 12403025.
J.C. acknowledge the support from National Key Research and Development Program of China No. 2023YFA1608100 and NSFC under grants No. 12273027.
F.-L.W. acknowledge the support from NSFC under grants No. 12073043.

Guoshoujing Telescope (the Large Sky Area Multi-Object Fiber Spectroscopic
Telescope LAMOST) is a National Major Scientific Project built by the Chinese
Academy of Sciences. Funding for the project has been provided by the National
Development and Reform Commission. LAMOST is operated and managed by the
National Astronomical Observatories, Chinese Academy of Sciences. 

Funding for the Sloan Digital Sky Survey (SDSS) has been provided by the
Alfred P. Sloan Foundation, the Participating Institutions, the National
Aeronautics and Space Administration, the National Science Foundation, the
U.S. Department of Energy, the Japanese Monbukagakusho, and the Max Planck
Society. The SDSS Web site is http://www.sdss.org/.

The SDSS is managed by the Astrophysical Research Consortium (ARC) for the
Participating Institutions. The Participating Institutions are The University
of Chicago, Fermilab, the Institute for Advanced Study, the Japan Participation
Group, The Johns Hopkins University, Los Alamos National Laboratory, the
Max-Planck-Institute for Astronomy (MPIA), the Max-Planck-Institute for 
Astrophysics (MPA), New Mexico State University, University of Pittsburgh,
Princeton University, the United States Naval Observatory, and the University
of Washington.

This work has made use of data from the European Space Agency (ESA) mission
{\it Gaia} (\url{https://www.cosmos.esa.int/gaia}), processed by the {\it Gaia}
Data Processing and Analysis Consortium (DPAC,
\url{https://www.cosmos.esa.int/web/gaia/dpac/consortium}). Funding for the DPAC
has been provided by national institutions, in particular the institutions
participating in the {\it Gaia} Multilateral Agreement\end{acknowledgments}

%

\vspace{5mm}


\software{emcee \citep{emcee}, 
		corner \citep{corner}
		MGEFIT \citep{capp02}
               JAMPY \citep{capp08, capp20}, 
          	Gala \citep{gala}}



%

\clearpage

\clearpage

\clearpage

\clearpage

\clearpage

\clearpage

\clearpage
%
%

\clearpage

\clearpage

\clearpage

\clearpage
\begin{figure}
	\gridline{\fig{KG_sigma_R_all_mod_multiPops.png}{0.25\textwidth}{}
		     \fig{KG_sigma_z_all_mod_multiPops.png}{0.25\textwidth}{}
	            \fig{KG_sigma_phi_all_mod_multiPops.png}{0.25\textwidth}{}
	            \fig{KG_CovRz_all_mod_multiPops.png}{0.25\textwidth}{}}
    	\caption{Similar to Fig.~\ref{fig:vd_compare_kg_mp},  but for the BHB samples.}
\label{fig:vd_compare_bhb_mp}
\end{figure}

\clearpage
\begin{figure}
	\gridline{\fig{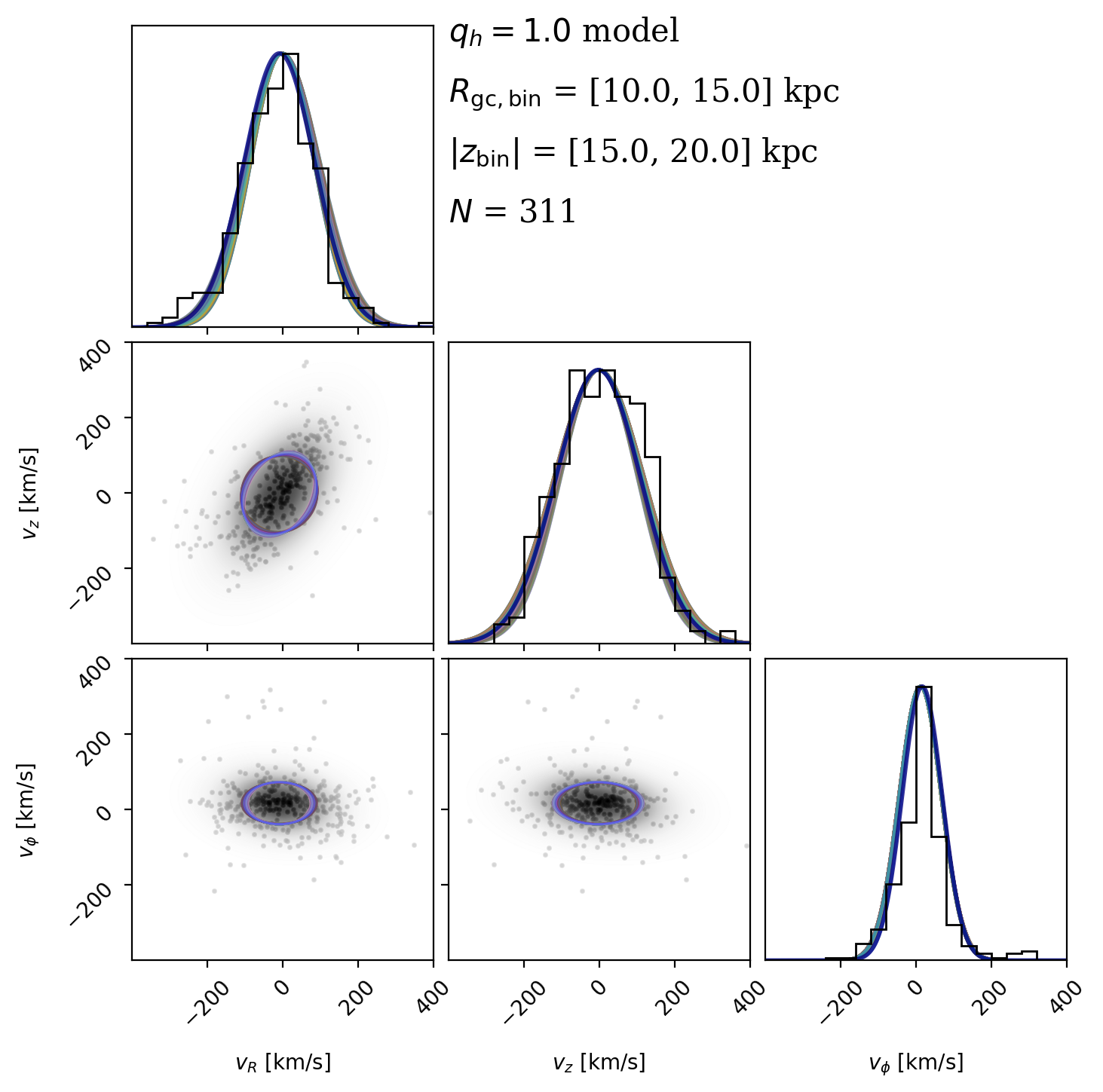}{0.3\textwidth}{}
	             \fig{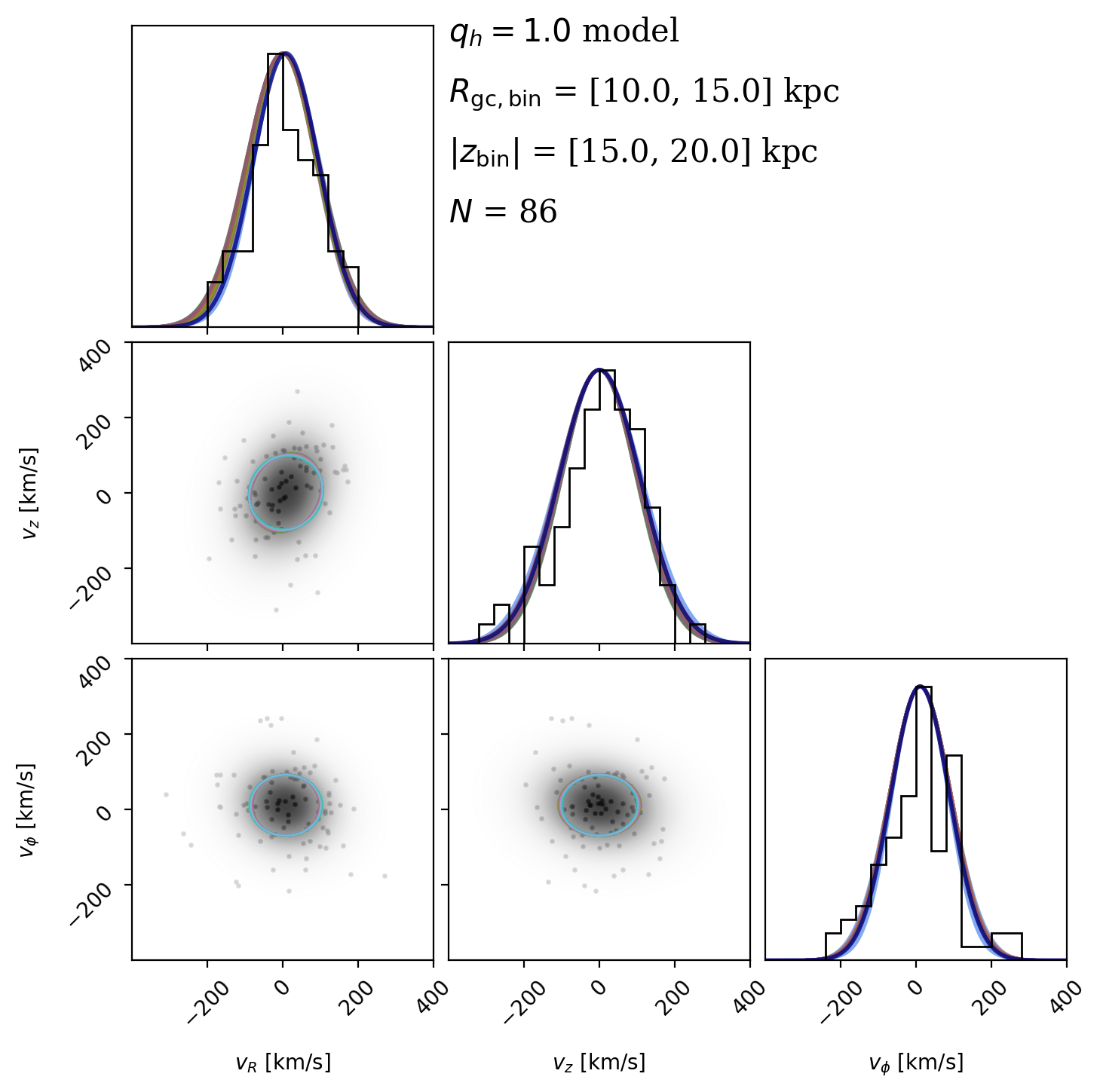}{0.3\textwidth}{}}
	\gridline{\fig{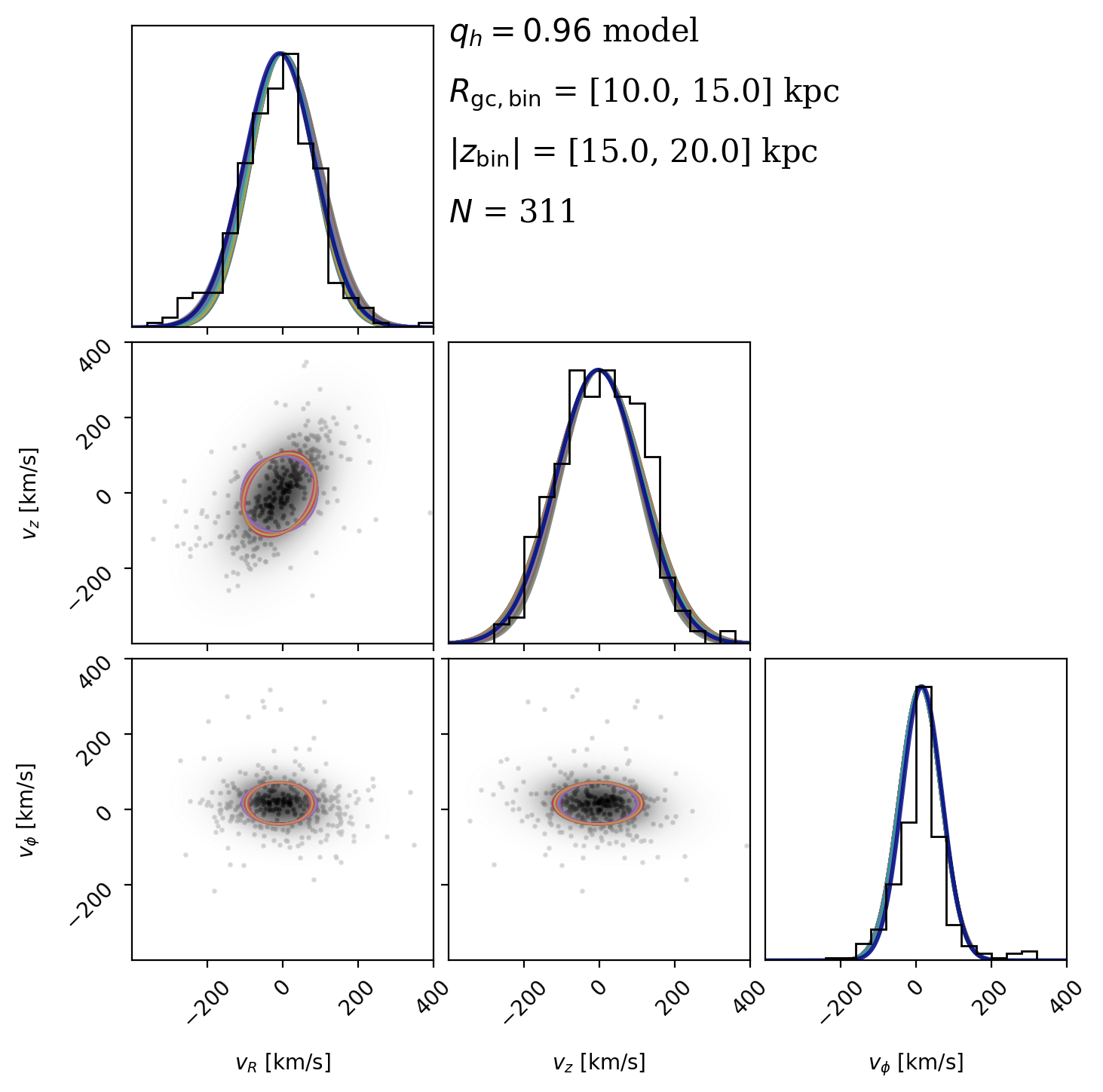}{0.3\textwidth}{}
	             \fig{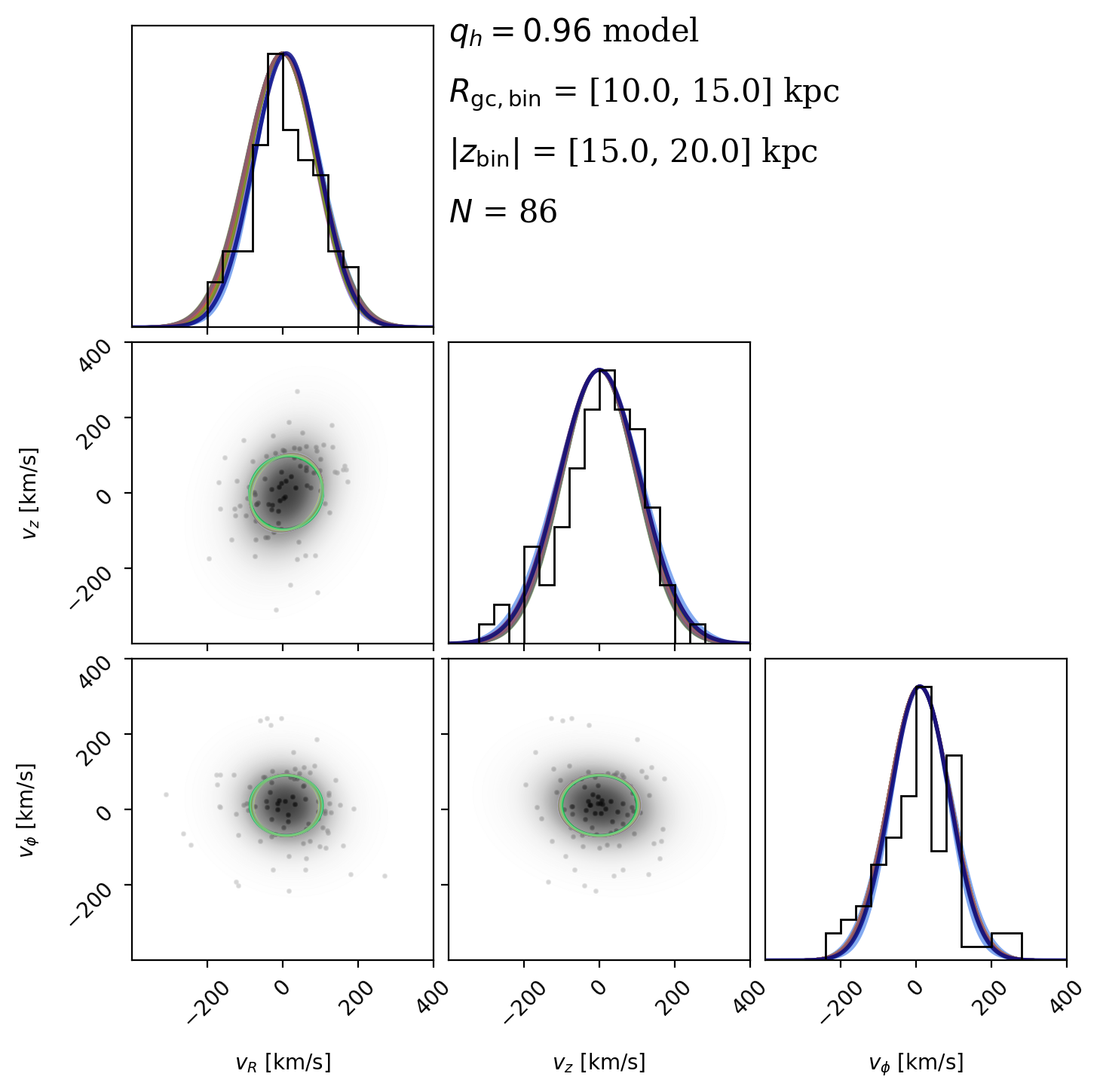}{0.3\textwidth}{}} 
	\gridline{\fig{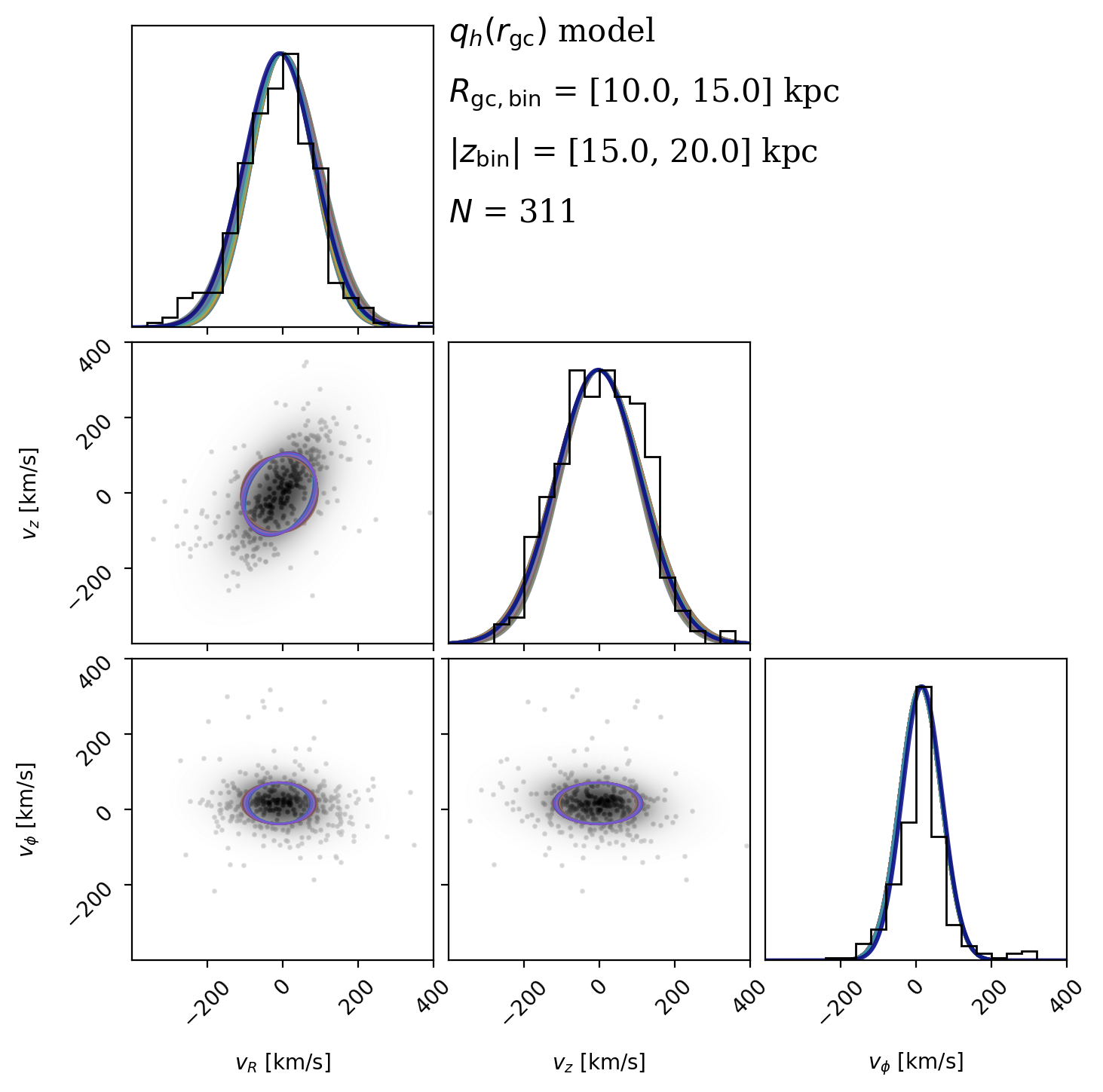}{0.3\textwidth}{KG}
	             \fig{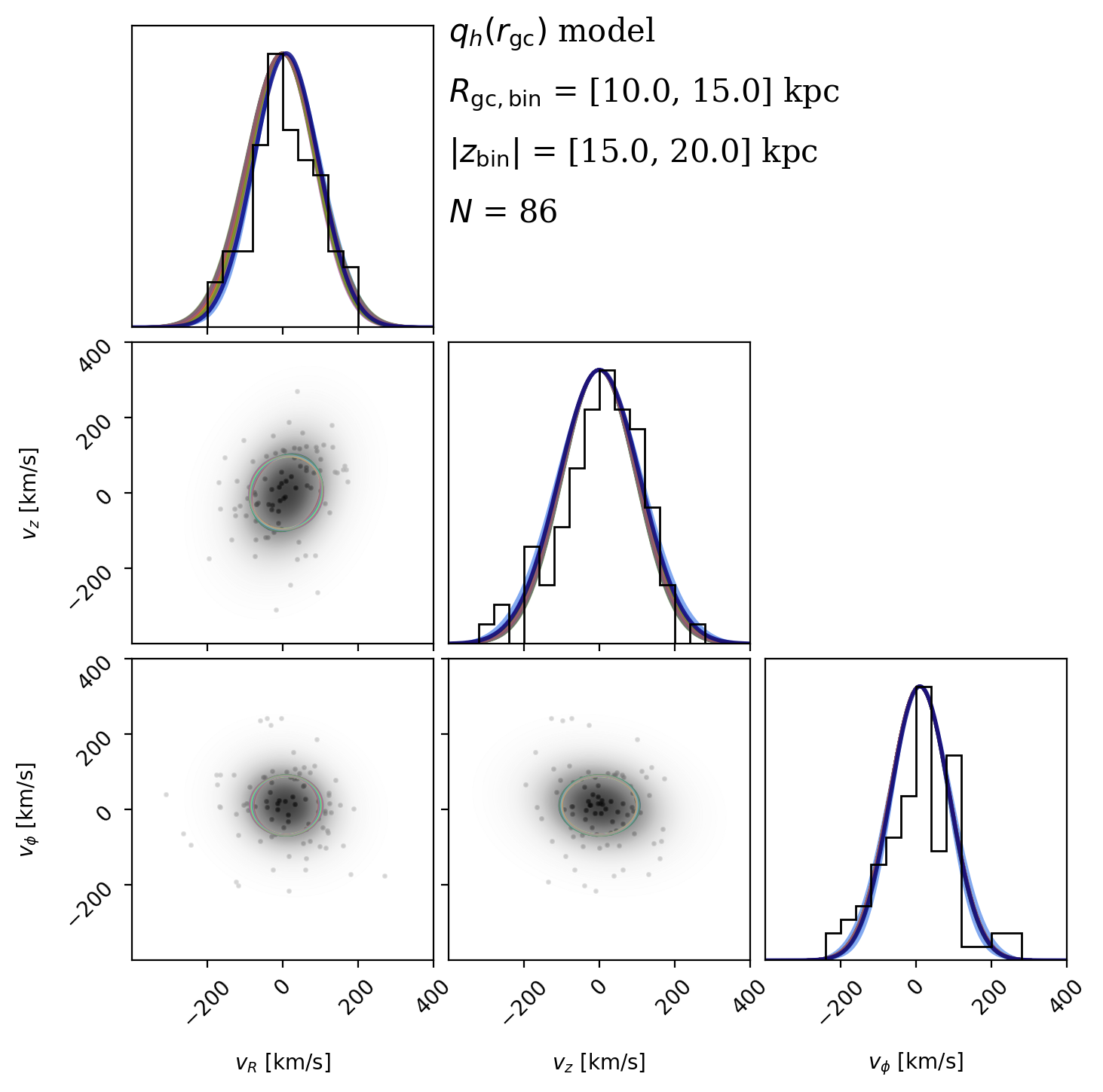}{0.3\textwidth}{BHB}}                                   
    	\caption{Example fits of fits of the components of velocity-dispersion tensors, shown for the two tracers in the range of $R \in [10., 15]$~kpc and
		      $|z| \in [15, 20]$~kpc. The models, from top to bottom, are $q_h=1$, $q_h = 0.96$, and $q_h(\rgce)$ ones, while the samples, from left to right,
		      are for KG and BHB. In each plot along the diagonal, the histograms represent the observed velocity distribution, with the blue solid line
		      showing the intrinsic velocity distribution accounting for the observed errors. The colored solid lines indicate the velocity distributions
		      predicted by different models for each sample within the $R-|z|$ cell. The off-diagonal plots represent the 2D projections of the velocity
		      ellipsoid in different velocity components. The black dots show the sample distribution in this velocity space, and the colored solid lines
		      show the projections of the velocity ellipsoids predicted by different models for each sample within the $R-|z|$ cell.}
\label{fig:vd_examples}
\end{figure}	              

\clearpage
\begin{figure}
	\gridline{\fig{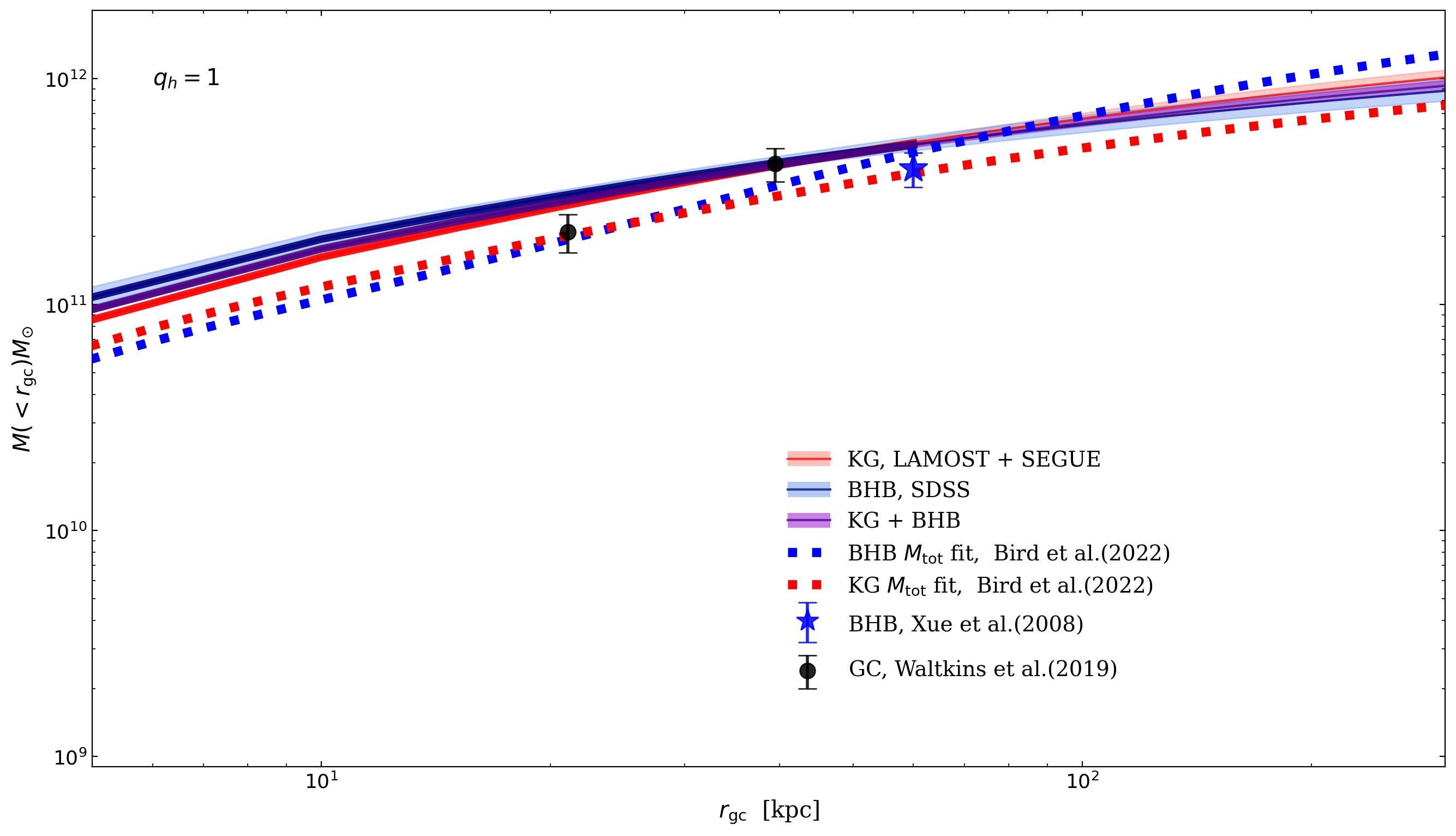}{0.5\textwidth}{}}
	\gridline{\fig{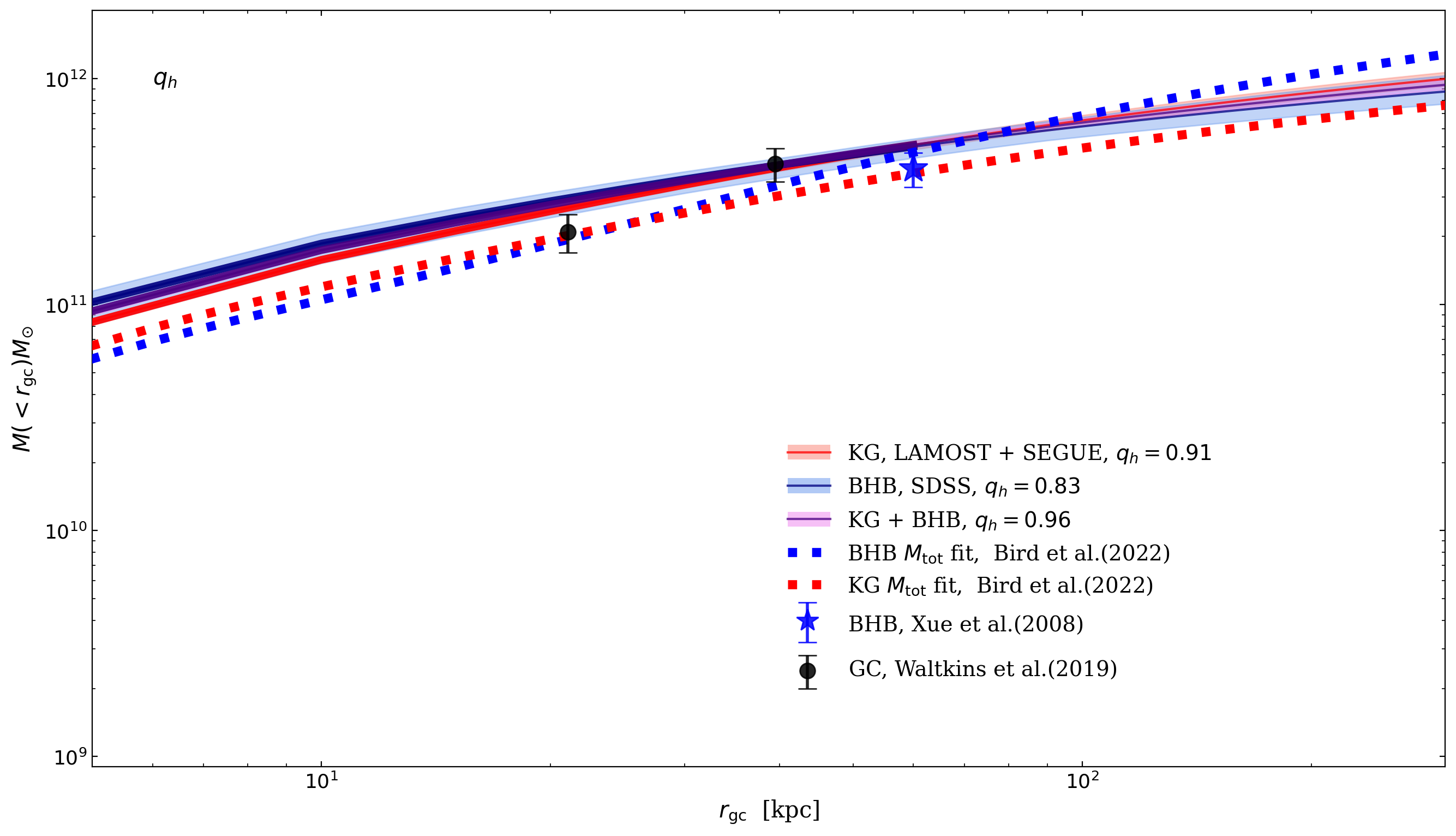}{0.5\textwidth}{}}
	\gridline{\fig{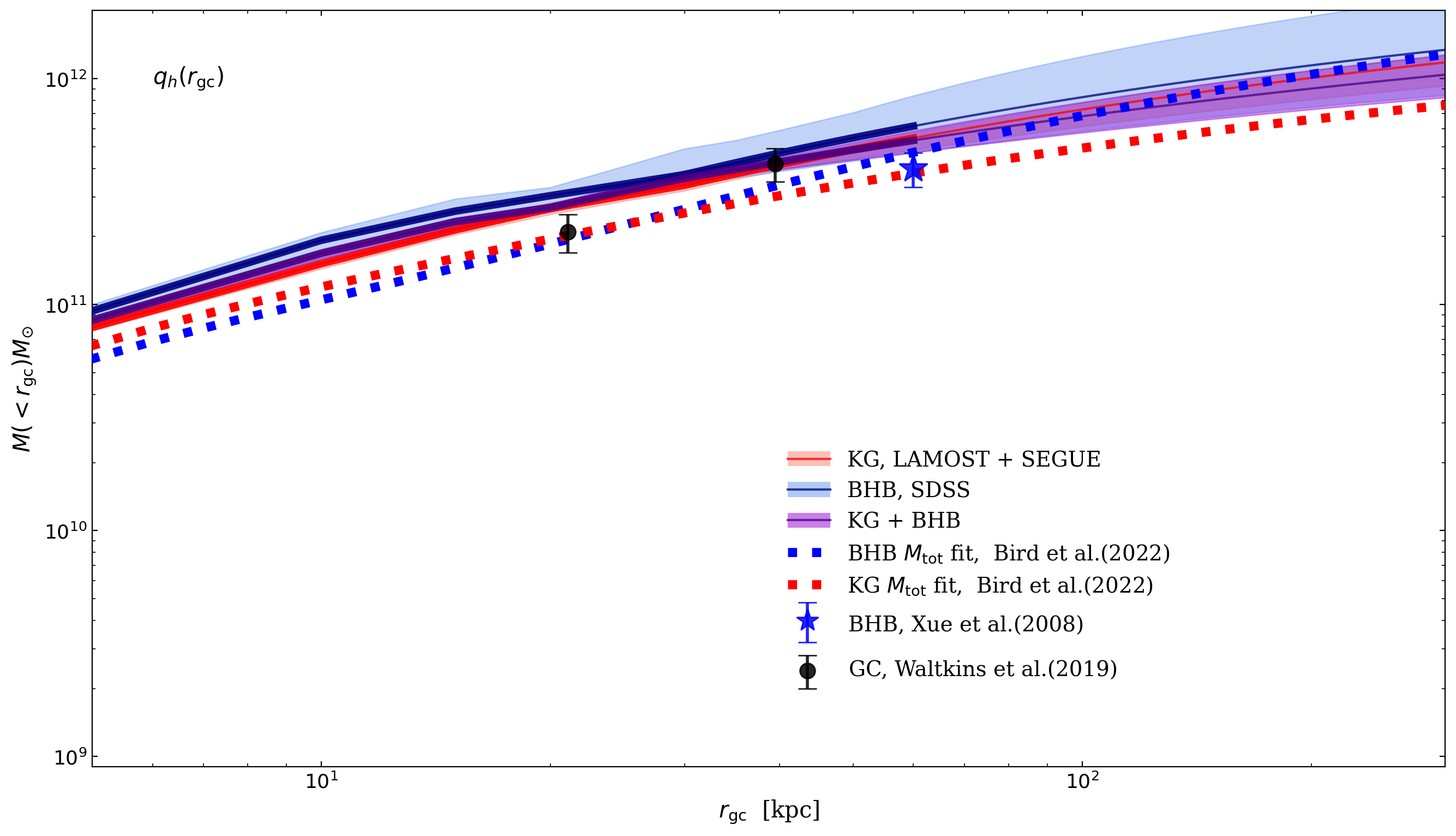}{0.5\textwidth}{}}                                            
    	\caption{ The enclosed mass calculated from the three models, including baryonic matter terms. The
	             red, blue, and purple solid lines represent the results obtained from fitting KG only, BHB only, and both tracers simultaneously, respectively.
	             The shaded regions correspond to the parameter sampling of the last 200 steps in the MCMC chain. The bold solid lines indicate that 90\% of the
	             data is covered within this distance range. The enclosed mass results from other literature studies are also shown from a comparison.}
\label{fig:rho_mass}
\end{figure}

\clearpage
\begin{figure}
	\gridline{\fig{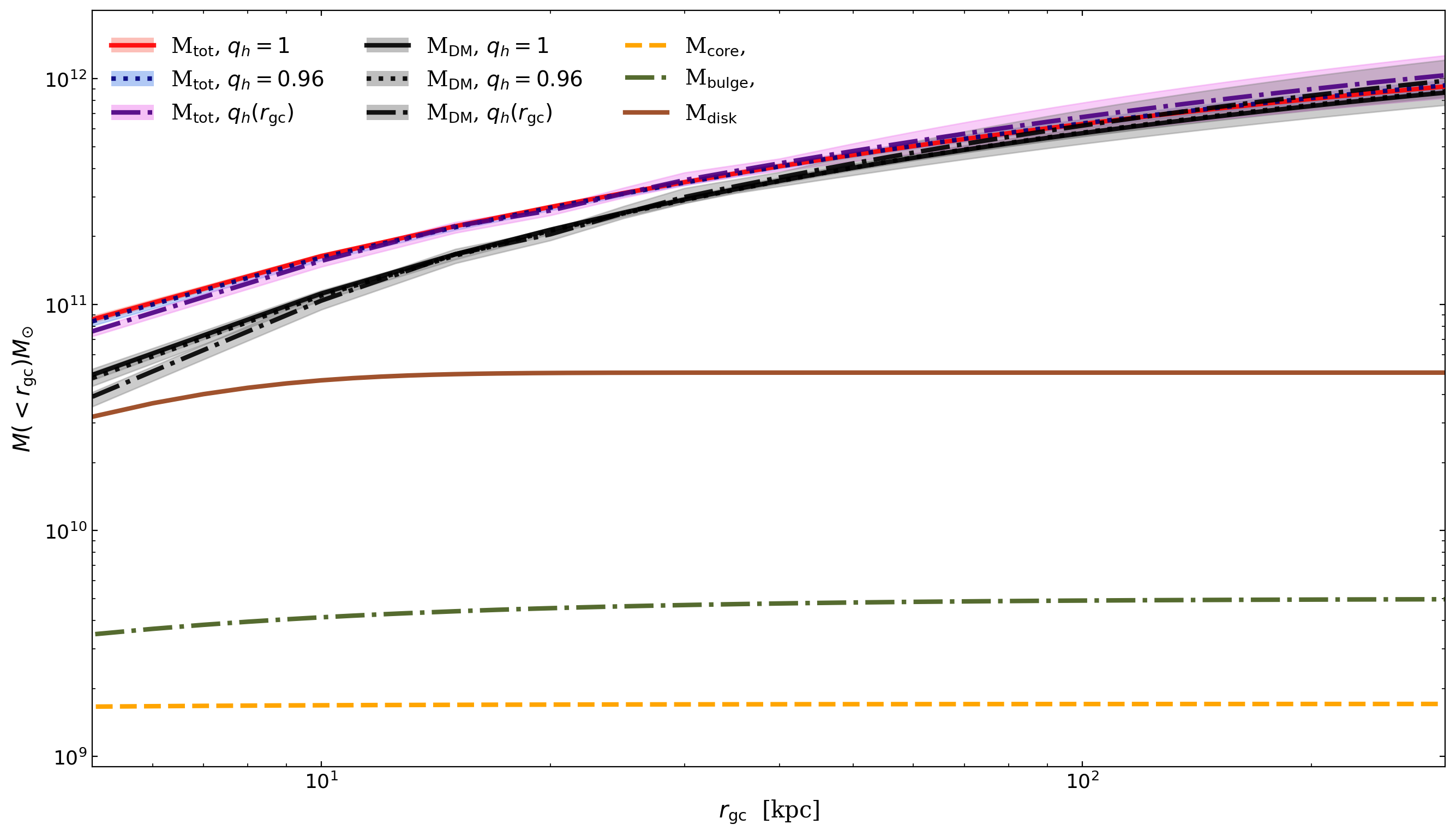}{0.6\textwidth}{}}
	\gridline{\fig{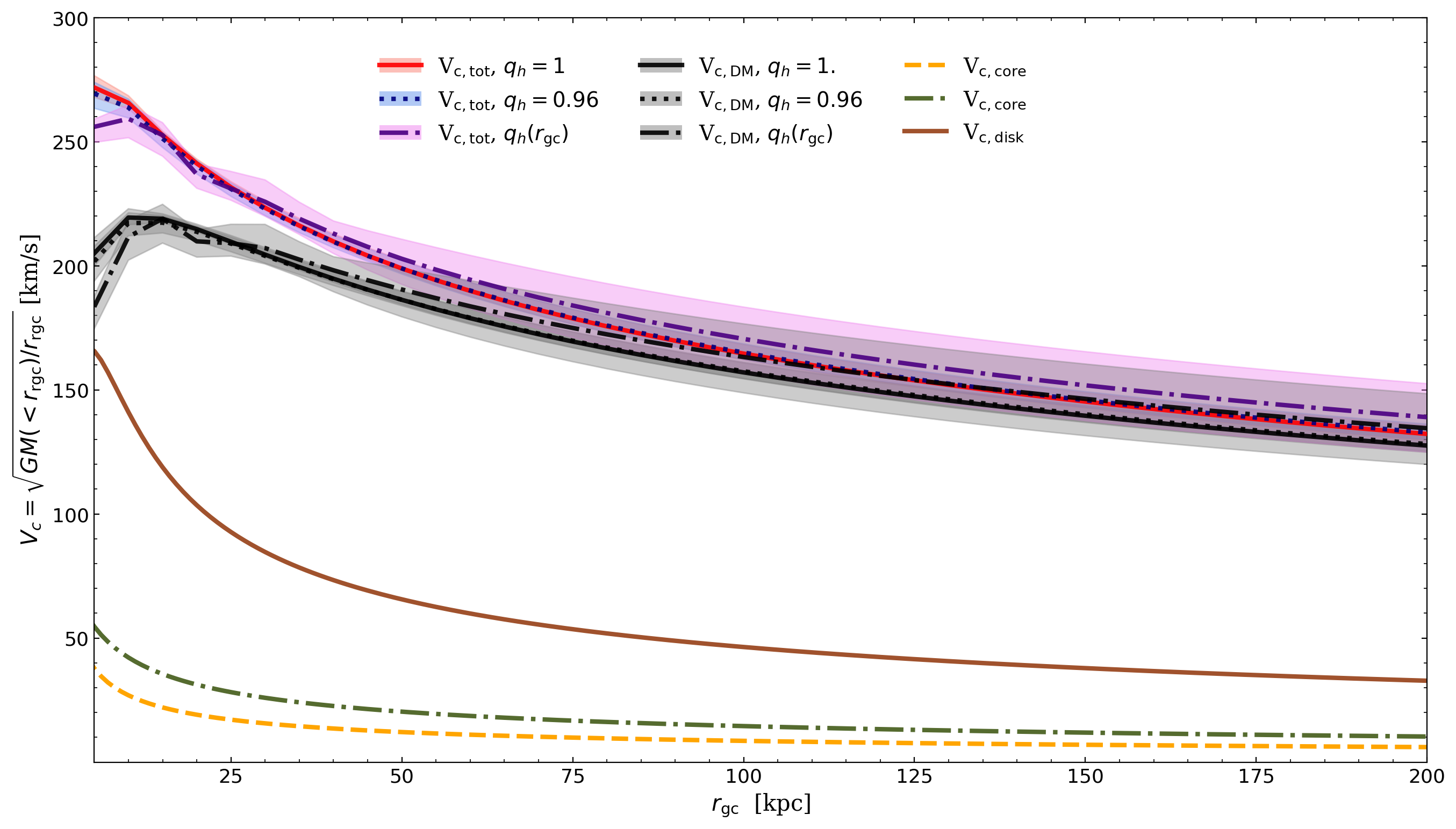}{0.61\textwidth}{}}                       
    	\caption{The upper panel is similar to Fig.~\ref{fig:rho_mass}, but for comparison of different dark matter density profile models.
	             While the lower one is the circular velocities under the models. Here we used the results from simultaneously fitting as an example. 
	             In the both of the plots, the contributions from different components are also shown.}
\label{fig:model_compare}
\end{figure}

\clearpage
\begin{figure}
\centering
	\includegraphics[width=0.8\textwidth]{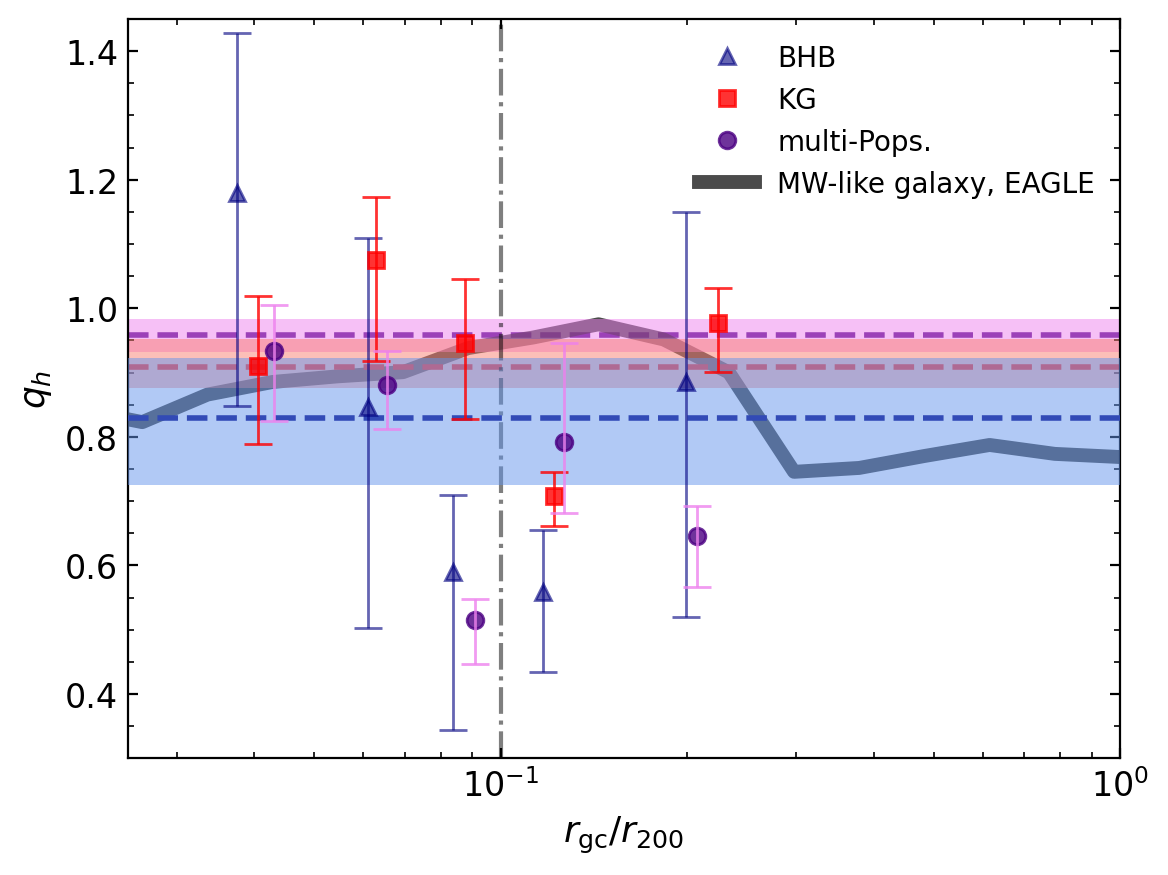}                                       
    	\caption{Comparison of fitted $q_h$ from different models. Dashed lines and correspond color bands are results and the $1-\sigma$ fitting errors
	            from $q_h$ model, while scatter points show the results of the $q_h(\rgce)$ model. The red, blue, and purple solid lines represent the
	            results obtained from fitting KG only, BHB only, and both tracers simultaneously, respectively. Meanwhile, the axial ratio of a MW analogue in the EAGLE simulation
	            \citep{shao21} are shown in black thick line. The inner/outer halo is separated by a dashed-dot line.}
\label{fig:q_compare}
\end{figure}

\clearpage
\begin{table}[!htb]
\centering
\caption{Best fitted parameters, in the case of broken power law number density and $q_h = 1$.}
\label{tab:best_fit_q1_bpl}
\begin{tabular}{ccrrr}
  \hline
  \multicolumn{2}{c}{Fitted Parameters} & KG only & BHB only & KG + BHB \\
  \hline
  \multicolumn{2}{c}{$\log \rho_s$ [\Msunpc]}     & $-1.156^{+0.047}_{-0.044}$  & $-0.539^{+0.130}_{-0.133}$ & $-0.861^{+0.042}_{-0.047}$ \\
  \multicolumn{2}{c}{$r_s$ [kpc]}                         & $7.319^{+0.351}_{-0.392}$  & $4.082^{+0.504}_{-0.572}$ & $5.482^{+0.252}_{-0.250}$ \\
  \hline
  \multirow{2}{*}{KG}   & $\beta$                        & $0.762^{+0.004}_{-0.003}$  & $...$                                  & $0.770^{+0.004}_{-0.003}$ \\
                                & $\gamma$                    & $0.803^{+0.003}_{-0.003}$  & $...$                                  & $0.819^{+0.003}_{-0.003}$ \\
  \hline
  \multirow{2}{*}{BHB} & $\beta$                         & $...$            & $0.581^{+0.018}_{-0.017}$ & $0.567^{+0.006}_{-0.006}$ \\
                                & $\gamma$                     & $...$            & $0.553^{+0.020}_{-0.019}$ & $0.516^{+0.007}_{-0.007}$ \\
   \hline   
   \multicolumn{2}{c}{BIC}  & 784680 & 99755 & 1482987 \\
   \hline                      
   \multicolumn{5}{c}{Calculations from the Fitted Parameters} \\
   \hline
   \multicolumn{2}{c}{$\rho_{\rm dm} (R_{\odot}, 0)$ [\Msunpc]} & $0.0134^{+0.0003}_{-0.0003}$ & $0.0153^{+0.0010}_{-0.0008}$ & $0.0142^{+0.0002}_{-0.0001}$ \\
   \multicolumn{2}{c}{$r_{200}$ [kpc]} & $186^{+5}_{-5}$ & $178^{+9}_{-6}$ & $180^{+4}_{-3}$\\
   \multicolumn{2}{c}{$M_{200}$ [$\times 10^{12}~M_{\odot}$]} & $0.795^{+0.072}_{-0.060}$ & $0.696^{+0.107}_{-0.067}$ & $0.728^{+0.046}_{-0.034}$ \\
   \hline
\end{tabular}
\end{table}

\clearpage
\begin{table}[!htb]
\centering
\caption{Best fitted parameters, in the case of broken power law number density and $q_h$ is constant with \rgc.}
\label{tab:best_fit_q_bpl}
\begin{tabular}{ccrrr}
  \hline
  \multicolumn{2}{c}{Fitted Parameters} & KG only & BHB only & KG + BHB \\
  \hline
  \multicolumn{2}{c}{$\log \rho_s$ [\Msunpc]}     & $-1.202^{+0.045}_{-0.042}$  & $-0.643^{+0.197}_{-0.144}$ & $-0.906^{+0.058}_{-0.049}$ \\
  \multicolumn{2}{c}{$r_s$ [kpc]}                         & $7.821^{+0.389}_{-0.453}$  & $4.732^{+0.656}_{-1.131}$ & $5.801^{+0.301}_{-0.375}$ \\
  \multicolumn{2}{c}{$q_h$}                                & $0.909^{+0.033}_{-0.043}$   & $0.830^{+0.105}_{-0.093}$ & $0.959^{+0.027}_{-0.025}$ \\
  \hline
  \multirow{2}{*}{KG}   & $\beta$                        & $0.760^{+0.004}_{-0.004}$  & $...$                                  & $0.769^{+0.004}_{-0.004}$ \\
                                & $\gamma$                    & $0.803^{+0.003}_{-0.003}$  & $...$                                  & $0.818^{+0.003}_{-0.003}$ \\
  \hline
  \multirow{2}{*}{BHB} & $\beta$                         & $...$            & $0.579^{+0.018}_{-0.018}$ & $0.568^{+0.007}_{-0.006}$ \\
                                & $\gamma$                     & $...$            & $0.555^{+0.019}_{-0.019}$ & $0.515^{+0.007}_{-0.008}$ \\
   \hline
   \multicolumn{2}{c}{BIC}  & 784681 & 99762 & 1482998 \\
   \hline 
   \multicolumn{5}{c}{Calculations from the Fitted Parameters} \\
   \hline
   \multicolumn{2}{c}{$\rho_{\rm dm} (R_{\odot}, 0)$ [\Msunpc]} & $0.0138^{+0.0006}_{-0.0006}$ & $0.0169^{+0.0029}_{-0.0030}$ & $0.0146^{+0.0004}_{-0.0004}$ \\
   \multicolumn{2}{c}{$r_{200}$ [kpc]} & $184^{+5}_{-5}$ & $177^{+10}_{-8}$ & $181^{+4}_{-3}$\\
   \multicolumn{2}{c}{$M_{200}$ [$\times 10^{12}~M_{\odot}$]} & $0.779^{+0.063}_{-0.061}$ & $0.687^{+0.125}_{-0.092}$ & $0.736^{+0.046}_{-0.037}$ \\
   \hline                     
\end{tabular}
\end{table}

\clearpage
\begin{table}[!htb]
\centering
\caption{Best fitted parameters, in the case of  broken power law number density and $q_h$ is a step function of \rgc.}
\label{tab:best_fit_mq_bpl}
\begin{tabular}{ccrrr}
  \hline
  \multicolumn{2}{c}{Fitted Parameters} & KG only & BHB only & KG + BHB \\
  \hline
  \multicolumn{2}{c}{$\log \rho_s$ [\Msunpc]}     & $-1.411^{+0.029}_{-0.027}$ & $-1.300^{+0.091}_{-0.090}$  & $-1.350^{+0.033}_{-0.021}$ \\
  \multicolumn{2}{c}{$r_s$ [kpc]}                        & $9.875^{+0.267}_{-0.265}$ & $ 9.781^{+1.232}_{-0.786}$     & $9.997^{+0.085}_{-0.088}$ \\
  \multicolumn{2}{c}{$q_{h, 0}$}                          & $0.910^{+0.113}_{-0.120}$   & $  1.180^{+0.249}_{-0.332}$   & $ 0.934^{+0.072}_{-0.104}$ \\
  \multicolumn{2}{c}{$q_{h, 1}$}                          & $1.074^{+0.098}_{-0.157}$   & $ 0.846^{+0.263}_{-0.343}$  & $0.880^{+0.059}_{-0.069}$ \\
  \multicolumn{2}{c}{$q_{h, 2}$}                          & $0.945^{+0.098}_{-0.118}$  & $ 0.590^{+0.120}_{-0.246}$     & $0.516^{+0.032}_{-0.062}$ \\ 
  \multicolumn{2}{c}{$q_{h, 3}$}                          & $0.708^{+0.047}_{-0.038}$ & $ 0.559^{+0.098}_{-0.124}$   & $0.792^{+0.155}_{-0.1114}$ \\  
  \multicolumn{2}{c}{$q_{h, 4}$}                          & $0.977^{+0.054}_{-0.076}$ & $ 0.886^{+0.264}_{-0.366}$  & $0.645^{+0.046}_{-0.080}$ \\
  \hline
  \multirow{2}{*}{KG}   & $\beta$                        & $0.760^{+0.004}_{-0.004}$  & $...$           & $0.767^{+0.005}_{-0.004}$ \\
                                & $\gamma$                    & $0.802^{+0.004}_{-0.004}$  & $...$           & $0.821^{+0.003}_{-0.003}$ \\
  \hline
  \multirow{2}{*}{BHB} & $\beta$                        & $...$  & $0.577^{+0.020}_{-0.022}$           & $0.562^{+0.008}_{-0.009}$ \\
                                & $\gamma$                    & $...$  & $0.556^{+0.020}_{-0.023}$             & $0.519^{+0.010}_{-0.010}$  \\                              
  \hline
  \multicolumn{2}{c}{BIC}  & 784670 & 99790 & 1483077 \\
   \hline
   \multicolumn{5}{c}{Calculations from the Fitted Parameters} \\
   \hline
   \multicolumn{2}{c}{$\rho_{\rm dm} (R_{\odot}, 0)$ [\Msunpc]} & $0.0135^{+0.0020}_{-0.0021}$ & $0.0171^{+0.0052}_{-0.0052}$ & $0.0159^{+0.0017}_{-0.0029}$ \\
   \multicolumn{2}{c}{$r_{200}$ [kpc]} & $196^{+6}_{-17}$ & $207^{+49}_{-33}$ & $188^{+15}_{-15}$\\
   \multicolumn{2}{c}{$M_{200}$ [$\times 10^{12}~M_{\odot}$]} & $0.938^{+0.095}_{-0.222}$ & $1.099^{+0.973}_{-0.446}$ & $0.820^{+0.210}_{-0.186}$ \\
   \hline  
\end{tabular}
\end{table}


\bibliography{ms}{}
\bibliographystyle{aasjournal}



\end{document}